\documentclass[aps,floatfix,showpacs,nofootinbib,,amsmath,amssymb]{revtex4}

\usepackage{graphicx,bbm,epsfig}

\def\lta{~\mbox{\raisebox{-.6ex}{$\stackrel{<}{\sim}$}}~}
\def\gta{~\mbox{\raisebox{-.6ex}{$\stackrel{>}{\sim}$}}~}

\begin{document}

\begin{flushright}
{arXiv:0805.0273} \\
UMN--TH--2644/08\\
FTPI--MINN-08/13\\
\end{flushright}
\vspace{0.5cm}

\title{The nonperturbative decay of SUSY flat directions}

\author{A. Emir G\"umr\"uk\c{c}\"uo\u{g}lu$^1$, Keith A. Olive$^{1,2}$, Marco Peloso$^1$ and 
Matthew Sexton$^1$}

\address{$^1$ School of Physics and Astronomy,
University of Minnesota, Minneapolis, MN 55455, USA}
\address{$^2$ William I Fine Theoretical Physics Institute,
University of Minnesota, Minneapolis, MN 55455, USA}

\begin{abstract}
We compute the nonperturbative decay of supersymmetric flat directions due to their D-term potential.
Flat directions can develop large vacuum expectation values (vevs) during inflation, and, if they are long-lived, this can strongly affect the reheating and thermalization stages after the inflation. We study a generic system of two $U(1)$ or $SU(2)$ flat directions which are cosmologically evolving after inflation. After proper gauge fixing, we show that the excitations of the fields around this background can undergo exponential amplification, at the expense of the energy density of the flat directions. We compute this effect for several values of the masses and the initial vevs of the two flat directions, through a combination of analytical methods and extensive numerical simulations. For a wide range of parameters the flat directions decay within their first few rotations.
\end{abstract}


\maketitle


\section{Introduction}
\label{sec:introduction}

Flat directions are prevalent in the MSSM \cite{tony} and in its simplest extensions. 
Scalar fields can develop a large vev along a flat direction during inflation \cite{excitflat,eeno}, though in supergravity models, this typically requires a nonminimal Kahler potential \cite{drt,gmo,cgmo}. 
If large field values are excited, flat directions can have several important effects in cosmology \cite{enqmaz}. For example, in the presence of a non-vanishing baryon number violating operator, the flat direction may be associated with a finite baryon number density and could account for the 
production of the baryon asymmetry of the Universe through the Affleck--Dine mechanism \cite{ad,linde}. 
More recently, it has been argued that the flat directions may also affect the thermal history of the universe after inflaton decays \cite{am1}. There are two main effects. First there is the delay of the thermalization of the inflaton decay products (resulting in a low reheat temperature) due to the fact that the vev of the flat directions provides a large mass to gauge bosons, which in turn suppresses the rates of the processes needed for thermalization \cite{am1}.~\footnote{As pointed out in \cite{op}, this delay argument assumes that the vevs of the flat directions break all the gauge symmetries of the Standard Model. If this is not the case, some of the gauge bosons remain light, and thermalization may proceed through them.} The second effect is that the oscillations of the flat directions can come to dominate over the (relativistic) inflaton decay products \cite{eeno}, so that the relevant stage of reheating is associated with the decay of the flat directions, rather than that of the inflaton field.

For definiteness we assume that there is only a single mass scale associated with the inflaton potential, set by the COBE normalization \cite{cobe} to be about $m_\psi \sim 10^{-7} \, M_p \,$ (where $M_p$ is the Planck scale). We also assume that the inflaton couples to the visible sector only through gravity.
It was shown in \cite{op} that, under these circumstances, the delay of the thermalization of the inflaton decay products takes place only if the vev of the flat direction generated during inflation satisfies $\phi_0 \gta \alpha^{3/2} \, M_p^{5/2} m_{\phi} / m_\psi^{5/2} \,$, where $\alpha$ is a characteristic coupling strength between fermions and the gauge bosons (hence, of order of the gauge fine structure constant) and $m_\phi$ the mass of the flat direction. For $\alpha^2 \sim 10^{-3} \,$, and for $m_\phi \sim 100 \, {\rm GeV}$ (since the mass comes from soft SUSY breaking terms) this gives $\phi_0 \gta 10^{17} \, {\rm GeV} \,$. It was also shown that the energy density of the flat directions comes to dominate over that of the inflaton decay products for $\phi_0 \gta M_p^{4/3} \, m_\phi^{5/12} / m_\psi^{3/4} \sim 10^{17} \, {\rm GeV}$.

For our canonical choices of masses, the two conditions turn out to be the same. That is,  the vevs of the flat directions need to be very large to produce a delay in thermalization or for the flat directions to dominate the energy density. A limit on the vev of flat directions may be derived from higher-order superpotential terms of the form $\phi^n / M^{n-3} \,$, with $n > 3 \,$. Gauge invariance can protect MSSM flat directions and for each class of flat directions there is a minimum value of $n$ for which 
the flat direction is lifted \cite{tony}. No MSSM flat direction is protected beyond $n=9 \,$. 
Clearly, the fact that such terms are allowed by gauge invariance, does not imply that they are indeed in the superpotential, since their presence is model-dependent (they may for instance be forbidden by additional symmetries). These non-renormalizable terms, if present, can be used to determine the vev 
along a flat direction during inflation. Given sufficient time, a vev of order
$ \phi_0 \sim \left( m_\phi \, M^{n-3} \right)^{1 / (n-2)} \,$ is generated, though typical inflationary models
built on a single scale will produce a vev of order ${\phi_0}^2 \sim H^3 \tau \sim m_\psi M_P$ \cite{eeno}. 
In order to obtain a vev as large as $\vert \phi \vert \gta 10^{17} \, {\rm GeV}$, we must require that nonrenormalizable terms with low $n$ are absent. For $M = M_p$, and $m_\phi = 100 \, {\rm GeV} \,$, we find that all terms with $n < 11$ must be absent. It should be noted, however,  that loop corrections in
supergravity during inflation may lead to much larger vevs, particularly in models of the no-scale form
\cite{gmo,cgmo}.

Delayed thermalization and/or domination by flat direction requires that the flat directions
themselves are long-lived.
As pointed out in \cite{op} this appears to be the case if one only considers the 
perturbative decay of the flat directions. 
The perturbative decay rate of flat directions is suppressed by their large vev, $\Gamma_\phi \sim m_\phi^3 / \phi^2 \,$ \cite{ad}. For $\phi_0 \gta 10^{17} \, {\rm GeV} 
\,$, one finds indeed that the perturbative decay of the flat directions takes place after they dominate. More specifically, one finds $\Gamma_\phi \sim m_\phi^{5/3} / M_p^{2/3} \,$ \cite{op}. Since the typical timescale for an oscillation of a scalar field along a
flat direction prior to decay  is $m_\phi^{-1}$, one finds that flat directions perform about $\left( M_p / m_\phi \right)^{2/3} \sim 10^{10}$ oscillations before their perturbative decay. 

However, it may be possible that the decay of a flat direction vev is controlled by 
nonperturbative effects. This possibility was first discussed in \cite{Allahverdi:1999je,Postma:2003gc},
where it was assumed that the scalar field along a flat direction has an interaction of the type $\Delta V = g^2 \vert \phi \vert^2 \vert \chi \vert^2$ with another scalar field $\chi$ \footnote{Nonperturbative effects in the presence of supersymmetric Q-balls were discussed in \cite{dan}.}.  The cosmological evolution 
of the flat direction then results in a time dependent effective mass for $\chi$. 
In principle, this interaction could lead to a nonadiabatic variation of the frequency for $\chi$, namely $\dot{\omega}_\chi / \omega_\chi^2 > 1$, and, consequently, to the nonperturbative production of $\chi \,$ at the expense of the energy stored along the flat direction. Such an effect is often included for the inflaton field, leading to nonperturbative inflaton decay or preheating \cite{kls}. 
It was, however, found that this effect is absent in the case of flat directions. The reason being, that contrary to the case of  inflation, flat directions are complex fields that typically do not pass through $\vert \phi \vert =0$ during their ``oscillations''.~\footnote{This is due to terms which break the radial symmetry of the potential of the flat direction, and which therefore generate an angular motion. One can however envisage a situation in which such terms are absent, so that the motion is purely radial. In this case, preheating effects like parametric resonance \cite{kls} or instant preheating \cite {Felder:1998vq}
lead to the fast decay of the flat direction~\cite{Giudice:2008gu}.} Therefore, the frequency $\omega_\chi$ is always large enough to ensure that $\dot{\omega}_\chi / \omega_\chi^2 \ll 1$ at all times.

Although formally correct, this conclusion heavily relies on the assumed coupling $\Delta V = g^2 \vert \phi \vert^2 \vert \chi \vert^2$. However, the couplings of MSSM flat directions to other fields is much more complicated than this, and may give rise to nonperturbative effects not considered in \cite{Allahverdi:1999je,Postma:2003gc}. In particular, the D-term potential results in a large nondiagonal and time dependent mass matrix for several MSSM fields which depends on the (time dependent) background of the flat directions. Even if the eigenvalues of such a mass matrix are constant, or slowly evolving, this mass matrix ``rotates'' (in field space) due to the rotation of the flat directions in their complex plane. Such an effect can result in a very strong (exponential) production of MSSM fields, at the expense of the energy density stored in the flat directions, which therefore would decay on a much quicker timescale ($\mathcal{O}\left( 10 - 100 \right)$ rotations) than that suggested by the perturbative decay rate \cite{op}.

It was shown in \cite{op} that the non-perturbative decay of a flat direction vev requires the presence of two or more simultaneous flat directions. In the case of a single direction, the rotation of the flat direction in its complex plane can be absorbed by a gauge redefinition. This should not be seen as a serious problem, since it is easy to verify that only in exceptional cases does a single flat direction exclude all others from being excited. Therefore, if the conditions in the early universe allowed the formation of a large vev for a flat direction (sufficiently long phase of inflation, a nonminimal Kahler potential, and absence of the nonrenormalizable interactions of low orders), 
then one should conclude that a system of non-mutually exclusive flat directions developed large vevs.  For the case of multiple flat directions, it was
shown \cite{op}  that for several simple cases the number of degrees of freedom involved is sufficiently large that the mixing in field space remains even after all the gauge redundancy has been removed. However, the actual computation of the decay was performed only in toy models, characterized by a potential for the fields that has the same structure of the MSSM D-term potential: in such models, the flat directions experienced very rapid nonperturbative decay. The computation of the decay in complete models, with all the gauge interactions properly taken into account was postponed to a separate publication, and  that is the main goal of the present work. As we will see, the result of the detailed calculation presented here agrees with those of \cite{op}.

The paper is organized as follows. In Section~\ref{sec:models}, we introduce U(1) and SU(2) models with one and two flat directions, and we motivate why multiple flat direction models are necessary in order to undergo nonperturbative decay. In Section~\ref{sec:formalism} we outline the formalism for computing this decay, 
discuss under what conditions  a large effect can be expected. In Section~\ref{sec:decay} we perform a detailed analysis of the models introduced in Section~\ref{sec:models}. This study allows us to clarify why the nonperturbative production that we are studying is absent in the single flat direction case, but is instead present for multiple flat directions. In Section~\ref{sec:results} we present our numerical results for the particle production when two flat directions are excited. The computation is summarized in the discussion Section~\ref{sec:discussion}, where we also compare our results with others that have appeared in the literature after \cite{op}. The paper is concluded by a number of Appendices, in which we collect some of the  technical steps of our calculations.

%
%
\section{Modeling NonPerturbative Flat Direction Decay}
\label{sec:models}
In this section, we present the models considered in this work.  
These are specific models of gauged  U(1) and SU(2)  flat directions.
We derive the light degrees of freedom in the unitary gauge which can play a role in 
the non-perturbative decay of the fields excited along the flat directions.

We first consider a U(1) gauge theory with two fields of charge $+e$ and $-e$, with Lagrangian~\footnote{The normalization for the electric charge is chosen to match that of the SU(N) generators.}
\begin{eqnarray}
&& {\cal L} = \vert D \phi_1 \vert^2 + \vert D \phi_2 \vert^2 - \frac{1}{4} \, F^2 - V \left( \phi_1 ,\, \phi_2 \right)
\nonumber\\
&& D_\mu \, \phi_i = \left( \partial_\mu - \frac{i \, e}{2} q_i A_\mu \right) \phi_i
\label{lagrangian}
\end{eqnarray}
where $q_{1,2} = 1 ,\, -1 \,$. The dominant potential term, originating from a D-term, is
\begin{equation}
V_{D_1} = \frac{e^2}{8} \left( |\phi_1|^2 - |\phi_2|^2 \right)^2 \label{pot2}
\end{equation}
which admits the flat direction $\vert \phi_1 \vert = \vert \phi_2 \vert \,$. In the unitary gauge, this theory has $6$ degrees of freedom: the three polarizations of the gauge boson, two light modes, corresponding 
to the real and imaginary part of the excitations along the flat direction, and one massive Higgs. 

This decomposition is worked out in detail in appendix \ref{app:oneU1}.
We expand the scalar fields about some background value $\Phi$, for which the  
D-term potential (\ref{pot2}) vanishes: $\phi_1 = \Phi + \delta \phi_1$ and $\phi_2 = \Phi + \delta \phi_2$. Inserting these expansions in the potential (\ref{pot2}), one finds that the light modes are defined by $\delta \phi_1 + \delta \phi_2$. These fields can be rotated into the real and the imaginary excitations of the flat direction itself. They are decoupled from the other fluctuations at the linearized level, and their eigenfrequencies vary only adiabatically with time. Therefore, such fields do no undergo parametric resonance.

The transverse components of the vector fields are also decoupled at the linearized level. On the contrary, the longitudinal component is coupled to the Higgs field. However, as we show below, this coupling is strongly suppressed by the smallness of the flat direction mass (with respect to its amplitude) and can be neglected in the physically relevant cases. Therefore, all four of the massive fluctuations can be thought as decoupled from each other. Their mass terms are of the form
\begin{equation}
V \supset - \frac{1}{2} e^2 \vert \Phi \vert^2 \vert \delta_H \vert^2 +  \frac{1}{2} e^2 \vert \Phi \vert^2 A_\mu A^\mu
\label{pot1}
\end{equation}
where the Higgs fluctuation $\delta_H$ is obtained from the combination $\delta \phi_1 - \delta \phi_2$ (see below). The equality of the mass of the Higgs and of the gauge boson is due to the fact that the model is chosen to reproduce the bosonic sector of a supersymmetric theory. Actually, the masses of these fields receive additional contributions from the soft supersymmetry breaking mass terms $m^2 \vert \phi_i \vert^2$ of the two scalars, which are however much smaller than the amplitude $\vert \Phi \vert$ and can be ignored for now. These terms break the degeneracy between the masses of the gauge boson and the Higgs.

The potential (\ref{pot1}) leads to the time dependent effective mass of $\delta_H$ and $A_\mu$
\begin{equation}
m_{\rm eff}^2 = e^2 \vert \Phi \left( t \right) \vert^2
\label{mass1}
\end{equation}
In analogy to what has been computed for inflationary preheating \cite{kls}, one can investigate if the time variation of the mass is quick enough to lead to nonperturbative particle production. This nonperturbative decay of the background fields takes place whenever the frequency of the quanta  varies nonadiabatically, $\dot{\omega} \gtrsim \omega^2 \,$. For relatively small momenta $p$, one finds $\omega^2 = p^2 + m_{\rm eff}^2 \simeq e^2 \vert \Phi \left( t \right) \vert^2 \,$, and thus
\begin{equation}
\frac{\dot{\omega}}{\omega^2} \sim \frac{m}{|\Phi|}
\label{resultpot1}
\end{equation}
using $\vert \dot{\Phi} \vert \simeq m \, \vert \Phi \vert$, since this is the scale that governs the dynamics of the background. Since, $m$ is of order the electroweak scale, while $\Phi$ is close to the Planck scale, it is difficult to construct a situation in which this quantity becomes near unity.  Based on this consideration~\cite{Allahverdi:1999je,Postma:2003gc}, it was concluded~\cite{am1} that the flat direction decays only perturbatively, leading to a suppressed reheating temperature and a solution to the gravitino problem.
However, this conclusion strongly depends on the coupling~(\ref{pot1}) assumed in these analyses, and in the resulting mass term~(\ref{mass1}).  In several concrete cases, the coupling is more complicated.   

This is indeed the case if more flat directions are present. To see this, we  add two more fields to the Lagrangian also of charge $+e$ and $-e$ (other choices are possible as well),  and write the D-term contribution to the potential,
\begin{equation}
V_{D_2} = \frac{e^2}{8} \left( |\phi_1|^2 - |\phi_2|^2 + |\phi_3|^2 - |\phi_4|^2 \right)^2 \label{pot3}
\end{equation}
We discuss the case in which two flat directions are present in this model (notice that three flat directions are also possible). There are ten real modes around these directions (coming from the four complex scalars and the gauge boson);  in the unitary gauge, they can be decomposed as follows (see appendix \ref{app:twoU1} for the detailed computation):  three are associated with the massive vector, four provide shifts to the two flat direction vevs, and the three remaining degrees of freedom (one heavy and two light ones) are coupled to each other in a Higgs mass matrix. The longitudinal component of the vector field is also coupled to these fields, so this is actually a system of four coupled fields; however - analogously to what happens for the single flat direction case - this coupling turns out to be negligible in the regime of small masses vs. amplitudes that we are interested in.

Denoting the flat direction vevs by $\Phi$ and $\tilde\Phi$, the Higgs mass matrix possesses a heavy part proportional to $e^2\Phi^2$ (and $e^2\tilde\Phi^2$) which originates from the D-term~(\ref{pot3}), and a light part composed of soft supersymmetry breaking masses $m^2_{\phi_i}$ as well as terms of order $\vert\dot\Phi / \Phi\vert^2$ and $\vert\dot{\tilde\Phi} / \tilde\Phi\vert^2$ which appear from gauge fixing and from making the kinetic terms canonical.  The light terms also come with time dependent prefactors such as $\frac{|\Phi|}{\sqrt{|\Phi|^2+|\tilde\Phi|^2}}$ and $\frac{|\tilde\Phi|}{\sqrt{|\Phi|^2+|\tilde\Phi|^2}}$.   The resulting time dependence of the Higgs mass matrix in general cannot be removed by a field redefinition, and so all of the eigenmasses retain this time dependence.  The spectrum consists of one heavy state and two light states of order $m_{\phi_i}$ and in the limit of zero momentum, they all oscillate at the frequency $m_{\phi_i}$.  Thus, for momenta lower or comparable to the supersymmetry breaking masses, 
\begin{equation}
\frac{{\dot\omega}_{light}}{\omega_{light}^2} \sim 1
\label{adiabatic1}
\end{equation}
which suggests a nonadiabatic evolution of these light states.  Similarly, it can be shown that the corresponding eigenvectors  of these light states ${\hat C}_{light}$ will execute relatively fast rotation in the space they span,
\begin{equation}
\frac{\vert \dot{\hat C}_{light} \vert}{\omega_{light}} \sim 1
\label{adiabatic2}
\end{equation}
which can also lead to nonadiabatic evolution. The second effect was highlighted in \cite{op}, and it is discussed in detail in Sections~\ref{sec:formalism}~and~\ref{sec:discussion} below.  Both effects must be considered when there are multiple fields involved in the Higgs mass matrix.

This argument for nonadiabatic evolution relies upon the existence of light eigenstates in the Higgs mass matrix and also on their time dependence. The existence of these light states is not uncommon in the presence of multiple flat directions.  For example, in \cite{op} it was shown that the simultaneous presence of the two flat directions $LLE^c$ and $QLD^c$ results in a Higgs mass matrix with time dependent light states mixed with one heavy state.  It is expected that the results~(\ref{adiabatic1})~and~(\ref{adiabatic2}) will be valid also for this case.

It must be stressed that the rapid change of the light states is not guaranteed, and in particular it will be suppressed when the ratio $\vert\tilde\Phi\vert/\vert\Phi\vert$ is very large or very small.  In these limits, there is a suppression in the Higgs mass matrix via the above mentioned factors $\frac{|\Phi|}{\sqrt{|\Phi|^2 + |\tilde\Phi|^2}}$ and $\frac{|\tilde\Phi|}{\sqrt{|\Phi|^2+|\tilde\Phi|^2}}$.  A suppression of this kind was also noted in \cite{am2}, but a concrete calculation of the effects was not performed.  Our numerical studies indicate that even if the ratio of maximum values of the vevs is as large as $\sim 10^{+3}$, the ratio $\vert\tilde\Phi\vert/\vert\Phi\vert$ is still of order one during some short periods of time.  We find that even in cases such as this, there is sufficient nonadiabatic evolution during these short periods of time to result in the rapid decay of the flat direction. This effect will be demonstrated in Section~\ref{sec:results}.

In the remainder of this section, we discuss the background evolution as well as the symmetries of the background under changes in the initial conditions and parameters. For the two field case, we consider the potential
\begin{equation}
m^2(\vert\phi_1\vert^2 + \vert\phi_2\vert^2) + \lambda (\phi_1^2\phi_2^2 + \mbox{h.c.}) + \frac{e^2}{8}(\vert \phi_1\vert^2 - \vert \phi_2\vert^2)^2
\label{pot4} 
\end{equation}
where the last term is the D-term potential (\ref{pot2}), while $m^2$ is a soft supersymmetry breaking mass of order the electroweak scale. This potential admits the D-flat direction with vev assignments  $\langle\phi_1\rangle=\langle\phi_2\rangle = \Phi$. \footnote{Note, however, that taking $m_1\ne m_2$ will result in flat direction vevs with more complicated time evolution.} The quartic term proportional to $\lambda$ may be present due to one loop contributions to the effective potential and are assumed to have magnitude $\lambda\sim \frac{m^2}{\vert\Phi_0\vert^2}$ in analogy to the baryon number violating operators in the Affleck-Dine mechanism \cite{ad}.  Here, $\Phi_0$ is the flat direction vev at the time the flat direction begins its oscillation.   These terms provide an initial angular motion in the complex plane and become subdominant to the other terms in the action as the universe expands.  The vev subsequently spirals down slowly to the origin as
discussed in detail below.  Without the quartic terms, the dynamics of the flat direction vev would be a straight line trajectory in the complex plane through the origin.

The cosmological context for the evolution of the flat directions, is a Universe initially dominated
by the scalar field oscillations of the inflaton before it decays.  Thus initially
we assume a matter dominated Universe with the dominant contribution being in the inflaton energy density $\rho_\psi$. We then have the background equations of motion
\begin{eqnarray}
& \ddot\Phi + 3 \frac{\dot R}{R} \dot\Phi + m^2 \Phi + 2 \lambda {\Phi^*}^3 =0  \nonumber \\
& \left(\frac{\dot R}{R}\right)^2 = \frac{8 \, \pi }{3 \, M_p^2}\left( 2 |\dot\Phi|^2 + 2 m^2 |\Phi|^2 + \lambda (\Phi^4 + h.c. ) + \rho_\Psi\right)
\label{scalarbackground1}
\end{eqnarray}
where we have assumed a Friedmann-Robertson-Walker metric $ds^2 = dt^2-R^2d\mathbf{x}^2$ in which $R$ is the scale factor.  
As we will see, it will be convenient to express the flat direction in terms of two scaled real
scalar fields, 
\begin{equation}
\Phi=\frac{F}{2 \, R} e^{i\Sigma}
\end{equation}
The equations of motion may be rewritten using conformal time $\eta$ with $dt \equiv R \, d\eta$ as,
\begin{eqnarray}
&F''+\left(m^2 R^2- \frac{R''}{R}-\Sigma'^2 \right)F + \frac{\lambda}{2} \,F^3 \cos\left(4\,\Sigma\right) = 0\nonumber \\
&\left(F^2\,\Sigma'\right)' -\frac{\lambda}{2} \,F^4 \sin\left(4\,\Sigma\right)=0 \nonumber \\
& \frac{R''}{R}+ \left(\frac{R'}{R}\right)^2 = {4\pi R^2 \over M_P^2} (2 V(\Phi) + \rho_\psi) \label{scalarbackground2}
\label{eombck}
\end{eqnarray}
where a prime denotes a derivative with respect to the conformal time, and where the Friedmann equations were combined in a way so as to eliminate the kinetic terms for the scalar fields.  This form~(\ref{scalarbackground2}) of the background equations is used for our numerical analyses shown below.  

The background equations of motion possess two symmetries.  Given a solution $\Phi_A(t)$ to the equations~(\ref{scalarbackground1}) with initial value $\Phi_A(0) = \Phi_0$, and recalling that $\lambda \propto \frac{m^2}{|\Phi_0|^2}$, we see that $\gamma \, \Phi_A(t)$, with $\gamma$ a real constant, is also a solution to the equations of motion.  Thus, changing the initial value for the field, only changes the solution through a scale change.  The background equations have a second symmetry under a change of the soft mass $m\rightarrow \mu \, m$, with $\mu$ real. Now,  $\Phi_A(\mu \, t)$ is a solution to the new background equations of motion.  Thus changing the mass parameter in the equations only changes the timescale on which the solution evolves.  Both symmetries will be shown in Section~\ref{sec:discussion} to translate into approximate symmetries for the decay of the flat directions.~\footnote{For this argument, we neglect the subdominant contribution to the Friedmann equation (\ref{scalarbackground1}) from the energy density of the flat direction. The initial energy of the inflaton field is chosen so that the flat direction is initially at rest due to Hubble friction. The flat direction starts evolving when the Hubble parameter $\propto \rho_\psi^{1/2}$ drops below its mass. Starting with a higher value for $\rho_\psi$ simply amounts in waiting for a longer time before this evolution starts. Therefore, under the rescaling $m \rightarrow \mu \, m$, we also rescale $\rho_\psi \rightarrow \mu^2 \, \rho_\psi \,$.}

This model can be extended to the four fields model with potential
\begin{eqnarray}
&& m^2(\vert\phi_1\vert^2 + \vert\phi_2\vert^2) 
+ {\tilde m}^2 (\vert\phi_3\vert^2 + \vert\phi_4\vert^2)
+\lambda (\phi_1^2\phi_2^2 + \mbox{h.c.})
+ {\tilde \lambda} (\phi_3^2\phi_4^2 + \mbox{h.c.}) \nonumber \\
&& + \;\frac{e^2}{8}(\vert \phi_1\vert^2 - \vert \phi_2\vert^2 + \vert \phi_3\vert^2 - \vert \phi_4\vert^2)^2
\label{pot5} 
\end{eqnarray}
characterized by the D-term potential (\ref{pot3}), soft supersymmetry breaking terms, and quartic interactions. This model admits the two flat directions characterized by the vevs $\langle\phi_1\rangle=\langle\phi_2\rangle = \Phi \,,\; \langle\phi_3\rangle=\langle\phi_4\rangle = {\tilde \Phi} \,$. The background equations of motion are an immediate extension of~(\ref{scalarbackground2}); we simply have two copies of the first two equations in ~(\ref{scalarbackground2}) (one copy for the amplitude and the phase of $\Phi$, and one for the amplitude and the phase of ${\tilde \Phi}$), while the last equation
in ~(\ref{scalarbackground2}) only changes in the sense that the potential for both fields is present on the right hand side. Since $\lambda \propto \frac{m^2}{|\Phi_0|^2}$, and ${\tilde \lambda} \propto \frac{{\tilde m}^2}{|{\tilde \Phi}_0|^2} \,$, the system possesses the same symmetries of the previous case under the simultaneous rescalings $\Phi \rightarrow \gamma \, \Phi ,\, {\tilde \Phi} \rightarrow \gamma \, {\tilde \Phi}$
or $m \rightarrow \mu \, m ,\, {\tilde m} \rightarrow \mu \, {\tilde m} \,$.

In this work, we study the decay of the background flat direction(s) into the fields involved in the potentials (\ref{pot4}) and (\ref{pot5}). We do so by studying the linearized theory for the perturbations, and by studying whether the system of fluctuations undergoes nonperturbative production due to the coherent motions of the flat direction(s).~\footnote{We comment on the effect of the nonlinear interactions in the Discussion section~\ref{sec:discussion}.} This linearized computation can be extended to nonabelian flat directions of physical interest, as for instance SU(N) flat directions. The reason for this is that the differences due to the nonabelian structure only arise at higher than quadratic order in the gauge fields. Such interactions are not present at the linearized study of fluctuations, as long as the gauge fields involved do not have any background expectation value. 

For simplicity, we focus on the SU(2) case. Consider the model with potential
\begin{equation}
m^2(\vert\phi_1\vert^2 + \vert\phi_2\vert^2) + \lambda \left[ (\varepsilon_{ij} \phi_{1i}\phi_{2j})^2 + \mbox{h.c.}\right]
+ \frac{e^2}{2}\sum_a \left(\phi_1^* \tau^a \phi_1 + \phi_2^* \tau^a \phi_2\right)^2
\label{pot6} 
\end{equation}
where $i,j$ are SU(2) doublet indices, and the $\tau^a$ are the gauge generators $\tau^a=\frac{\sigma^a}2$ (where $\sigma^a$ are the Pauli matrices). This potential admits the flat direction
$\phi_1 = \left( \Phi ,\, 0 \right) \,,\; \phi_2 = \left( 0 ,\, \Phi \right) \,$. As we show in Appendix~\ref{app:SU2}, the physical modes around this background consist of the real and imaginary excitations of the flat direction, of three massive gauge bosons, and of three Higgses. The 
quadratic actions for the gauge bosons and the Higgses are three copies of those obtained in the U(1) case.

We can extend this model to contain four complex doublets, with potential
\begin{eqnarray}
&&  m^2(\vert\phi_1\vert^2 + \vert\phi_2\vert^2)
+ {\tilde m}^2(\vert\phi_3\vert^2 + \vert\phi_4\vert^2)
+ \lambda \left[ (\varepsilon_{ij} \phi_{1i}\phi_{2b})^2 + \mbox{h.c.}\right]
+ \tilde \lambda \left[ (\varepsilon_{ij} \phi_{3i}\phi_{4j})^2 + \mbox{h.c.}\right] \nonumber \\
&& +\; \frac{e^2}{2}\sum_a \left(\phi_1^* \tau^a \phi_1 + \phi_2^* \tau^a \phi_2 + \phi_3^* \tau^a \phi_3 + \phi_4^* \tau^a \phi_4  \right)^2
\label{pot7} 
\end{eqnarray}
We can now have two flat directions, with the vev assignments $\phi_1 = \left( \Phi ,\, 0 \right) \,,\; \phi_2 = \left( 0 ,\, \Phi \right) \,,\; \phi_3 = \left( {\tilde \Phi} ,\, 0 \right) \,,\; \phi_4 = \left( 0 ,\, {\tilde \Phi} \right)  \,$. Also in this case, beside the real and imaginary parts of the two flat directions, the physical modes around this background consist of three copies of those obtained in the U(1) case; namely, three massive gauge bosons, three Higgses, and six light fields. Each Higgs is coupled to one longitudinal boson and two light fields exactly as for two U(1) flat directions.

Due to this correspondence between the linearized U(1) and SU(2) (which can be generalized to SU(N)) systems, it is sufficient to explicitly perform the numerical computations only for the U(1) cases.


\section{Formalism for the Quantum Evolution of Coupled Scalar Fields}
\label{sec:formalism}
Here we outline the formalism of \cite{Nilles:2001fg} which is necessary to quantify the rate of the nonperturbative decay of flat directions.  We begin with a Lagrangian with an arbitrary number of real scalar fields $\Psi=\{\psi_1,\psi_2,...\psi_N\}$ and with a time dependent mass matrix $M^2(\eta)$,
\begin{equation}
\mathcal{L} = \frac12 \partial_\mu{\Psi}^T \partial^\mu\Psi - \frac12 \Psi^T M^2(\eta) \Psi 
\nonumber
\end{equation}
where we are using the conformal time $\eta$, and derivatives with respect to conformal time are again denoted with a prime.  One may diagonalize $M^2$ with a time dependent rotation matrix $C(\eta)$,
\begin{equation} 
C^T(\eta)M^2(\eta)C(\eta) = m_d^2(\eta) \;\;\; \mbox{diagonal}
\label{massdiagonal}
\end{equation}
The eigenfrequencies of the system are the elements of the diagonal matrix
\begin{equation}
\omega \equiv \sqrt{k^2 \mathbbm{1} + m_d^2}
\end{equation}
where $k$ is the momentum. The fields in the diagonal basis are 
\begin{equation} 
\tilde \Psi \equiv C^T \Psi
\nonumber
\end{equation}
so if the mass matrix is evolving slowly enough, the column vectors of C denoted by ${\hat C}_i$ will be the physical eigenstates of the system.
The equations for the quantum evolution of the state of this system were shown in \cite{Nilles:2001fg} to be
\begin{eqnarray}
& \alpha' = \left(-i \omega -I\right)\alpha + \left(\frac{\omega'}{2\omega}-J\right)\beta \nonumber \\
& \beta' = \left(i \omega-I\right) \beta + \left(\frac{\omega'}{2\omega} -J\right)\alpha
\label{bogequations}
\end{eqnarray}
where $\alpha$ and $\beta$ are matrices of Bogolyubov coefficients for the set of scalar fields, $\Psi$ and
\begin{equation} I,J = \frac12 \left(\sqrt{\omega} \Gamma \frac{1}{\sqrt{\omega}} \pm \frac{1}{\sqrt{\omega}}\Gamma \sqrt{\omega} \right) \;\;\;,\;\;\; \Gamma = C^T C' 
\label{defineIJGamma}
\end{equation}
Note that  $\Gamma$ and $I$ are antisymmetric matrices while $J$ is symmetric.
In addition, the Bogolyubov coefficients obey the constraints,
\begin{eqnarray}
& \alpha\alpha^\dagger -\beta^*\beta^T = \mathbbm{1}  \label{bogconstraint1}\\
& \alpha\beta^\dagger- \beta^*\alpha^T = 0 \label{bogconstraint2}
\end{eqnarray}
and the occupation number of the i'th state can be shown to be,
\begin{equation} 
n_i (\eta)= \left(\beta^* \beta^T\right)_{ii}   \;\;\;,\;\;\;\mbox{no summation on }i
\label{occupationnumber}
\end{equation}
Note that the Gamma matrix may be written in terms of the eigenstates ${\hat C}_i$ and its elements contain the information for the rate of change of the eigenstates,
\begin{eqnarray}
\Gamma_{ij} &=& \hat C_i \cdot \hat C_j'  \nonumber \\
\hat{C}_i' &=& \sum_j -\Gamma_{ij} \hat C_j \nonumber
\end{eqnarray} 

There are two qualitatively different parts to the evolution equations~(\ref{bogequations}); a ``unitary" part represented by the anti-Hermitian matrices $(\pm i\omega - I)$ and a ``nonunitary" part represented by the symmetric matrix $\left(\frac{\omega'}{2\omega}-J\right)$.  Whenever the nonunitary part vanishes, the total occupation number given by $\mbox{Tr}[\beta^*\beta^T]$ is conserved. However, the matrix $\beta^*\beta^T$ itself  is not constant in general indicating a conversion from one species of particles to another as a function of time.  The non-unitary part will boost (or contract) both $\alpha\alpha^\dagger$ and $\beta^*\beta^T$ in~(\ref{bogconstraint1}) while keeping the difference invariant, so this term will change the total occupation number of the system.  It will be convenient to rewrite the evolution equations~(\ref{bogequations}) by extracting factors of $\sqrt{\omega}$, 
\begin{eqnarray}
\alpha' &=& \sqrt{\omega}\left(-i  - \frac{1}{\sqrt{\omega}}I\frac{1}{\sqrt{\omega}}\right)\sqrt{\omega}\alpha 
+ \sqrt{\omega}\left(\frac{\omega'}{2\omega^2}-\frac{1}{\sqrt{\omega}}J\frac{1}{\sqrt{\omega}}\right)\sqrt{\omega}\beta 
\nonumber \\
\beta' &=& \sqrt{\omega}\left(i  - \frac{1}{\sqrt{\omega}}I\frac{1}{\sqrt{\omega}} \right) \sqrt{\omega}\beta 
+ \sqrt{\omega}\left(\frac{\omega'}{2\omega^2} -\frac{1}{\sqrt{\omega}}J\frac{1}{\sqrt{\omega}}\right)\sqrt{\omega}\alpha
\label{bogequations2}
\end{eqnarray}
The equations are now written entirely in terms of the following matrices,
\begin{equation}
\sqrt{\omega}\;\;\;,\;\;\;
\left(1  \pm i \frac{1}{\sqrt{\omega}}I\frac{1}{\sqrt{\omega}}\right) \;\;\;,\;\;\;
A \equiv \frac{\omega'}{\omega^2}-\frac{1}{\sqrt{\omega}} \, 2J \, \frac{1}{\sqrt{\omega}}
= \frac{\omega'}{\omega^2} - \left( \Gamma \frac{1}{\omega} - \frac{1}{\omega}\Gamma \right)
\end{equation}
One recognizes the diagonal part of $A$ as the adiabatic parameter from the single-field analysis.  The off-diagonal parts of this matrix may also be interpreted as adiabatic parameters as follows.  One expects nonadiabatic evolution when either the oscillation frequency of the physical state or the field composition of the state (its eigenvector) has changed faster than its period $1 / \omega$.  The latter condition on the field composition is clearly relevant only if there are two or more fields from which the state is composed. Non-adiabatic evolution occurs when the condition~(\ref{adiabatic2}) is satisfied, $|\hat C_i'| > \omega_i$ or more specifically when,
\begin{equation}
\sqrt{\sum_j \left(\frac{\Gamma_{ij}}{\omega_i}\right)^2} > 1
\label{adiabaticguess}
\end{equation}
This condition is not quite correct however, since in the case of degenerate eigenstates, the eigenvectors have no preferred directions in their subspace, and  so the rate of change of these eigenvectors in this subspace is not physically meaningful.  This feature is taken into account by the $A$ matrix (as well as the $J$ matrix) which has the form,
\begin{eqnarray}
A_{ii} &=& \frac{\omega_i'}{\omega_i^2} \;\;\;\;\;\;\;\;\;\;\;\;\;  \mbox{no summation} \nonumber \\
A_{ij}  &=& \Gamma_{ij}\left(\frac{1}{\omega_i}-\frac{1}{\omega_j} \right)\;\;\;,\;\;\; i\ne j\;\; \mbox{ and no summation}
\label{adiabaticmatrix}
\end{eqnarray}
and it handles both limiting cases,
\begin{eqnarray}
\lim_{\omega_i\rightarrow\omega_j} A_{ij} =0 \;\;\;,\;\;\;i\ne j  
\nonumber \\
\lim_{\omega_j >> \omega_i} A_{ij} = \left(\frac{\Gamma_{ij}}{\omega_i}\right)  \;\;\;,\;\;\;i\ne j 
\nonumber
\end{eqnarray}
so for nondegenerate frequencies the condition (\ref{adiabaticguess}) is asymptotically true.  We thus take the elements of $A$ as our adiabatic parameters.  If any one element of the matrix is greater than unity $|A_{ij}|>1$, we expect nonadiabatic evolution.  We note finally that large $A_{ij}$ are a \textit{necessary} condition for net production of quanta, but not a \textit{sufficient} condition.


\section{Nonperturbative Decay of Single or Multiple Flat Directions}
\label{sec:decay}

In this section, we start the study of the nonperturbative decay of one flat direction with potential (\ref{pot4}) and of two flat directions with potential (\ref{pot5}). The spectra of fields for the two cases are worked out in appendices \ref{app:oneU1} and \ref{app:twoU1}, respectively. The results of these computations are summarized in the two Subsections below. The study of the quadratic action for these modes allows one to understand whether the nonperturbative decay of the flat directions(s) takes place. The actual numerical computation is presented in the next section.

\subsection{Single flat direction}

The single flat direction case is characterized by the potential~(\ref{pot4}), and the background values
\begin{equation}
\langle \phi_1 \rangle = \langle \phi_2 \rangle = \frac{F}{2 R} e^{i\Sigma} \;\;\;,\;\;\; 
\langle A_\mu \rangle = 0
\end{equation}
The perturbations of these fields can be most usefully written as
\begin{equation}
\delta \phi_1 + \delta \phi_2 = \frac{1}{R} \left( \delta_r + i \, \delta_i \right) \;\;\;,\;\;\;
\delta \phi_1 - \delta \phi_2 = \frac{{\rm e}^{i \Sigma}}{R} \, \left( \delta_H + i \, \delta_G \right)
\end{equation}

The mode $\delta_G$ is the only one in this decomposition that varies under an infinitesimal gauge transformation (since the two fields have opposite U(1) charge), and it can be set to zero in the unitary gauge.

The two fields $\delta_r$ and $\delta_i$ are the physical excitations along the real and imaginary directions of the flat directions. They are decoupled from the other fluctuations at the linearized level. However, they are coupled to each other in their mass term, due to the quartic term in~(\ref{pot4}),
\begin{equation}
m_{\rm flat}^2 = \frac{1}{R^2} \left( \begin{array}{cc}
m^2 R^2 - \frac{R''}{R} + \frac{3 \, \lambda}{2} \, F^2 \, \cos \left( 2 \Sigma \right) &
- \frac{3 \, \lambda}{2} \, F^2 \, \sin \left( 2 \Sigma \right) \\
- \frac{3 \, \lambda}{2} \, F^2 \, \sin \left( 2 \Sigma \right) &
m^2 R^2 - \frac{R''}{R} - \frac{3 \, \lambda}{2} \, F^2 \, \cos \left( 2 \Sigma \right) 
\end{array} \right)
\label{massmatflat}
\end{equation}
corresponding to the eigenmasses
\begin{equation}
m_{1,2}^2 = m^2 - \frac{R''}{R^3} \pm \frac{3 \, \lambda}{2} \, \frac{F^2}{R^2}
\end{equation}
The time variation of these eigenmasses is adiabatic. Moreover, the off-diagonal quartic term, already subdominant at the start, quickly becomes negligible as the flat direction starts evolving. Consequently,  the nonperturbative production of these fields is negligible as well.

The remaining mode $\delta_H$ is the Higgs field of the model. It is coupled to the longitudinal component of the gauge boson. The quadratic action for these two modes (after integrating out the non-dynamical fluctuation $A_0$) is given in (\ref{az1U1LH0s}). We study this action in the phenomenologically relevant case in which the masses of the flat directions are much smaller than their amplitudes. More accurately, we work in the limit of
\begin{equation}
\left\{ m \, R ,\,  \Sigma' ,\, \frac{F'}{F} \right\} \ll F
\label{seriesapproximation1}
\end{equation}
For bookkeeping, in the following we denote by $\epsilon$ the ratio between any of the terms on the left hand side of (\ref{seriesapproximation1}) and the term on the right hand side in this expression. Moreover, we disregard the terms proportional to $\lambda$ and $R''/R$ which become negligible as the flat direction starts evolving. 
In this limit, the masses of these two fields are
\begin{equation}
m_{\rm Higgs}^2 = \frac{e^2 \, F^2}{4 \, R^2} + \left[ m^2 + \frac{3 \, \Sigma'^2}{R^2} \right] + \mathcal{O} 
\left( \epsilon^4 \right) \;\;\;,\;\;\; m_{\rm long.}^2 = \frac{e^2 \, F^2}{4 \, R^2} + \mathcal{O} \left( \epsilon^4 \right) 
\label{masshL1}
\end{equation}
where the terms in square parenthesis are of order $\epsilon^2 \,$. The time variation of these masses is adiabatic, and, therefore, does not lead to particle production. Also the mixing between the two modes is suppressed by higher powers of $\epsilon \,$, and therefore does not lead to any relevant physical effect.

Finally, the spectrum of the fluctuations of this model also contains the two transverse polarizations of the gauge field, which are decoupled from the other modes (and from each other) at the linearized level.
Their masses coincide with the longitudinal component given in~(\ref{masshL1}).

Therefore, we conclude that a single flat direction does not experience nonperturbative decay, in agreement with what was argued in~\cite{op}.

\subsection{Multiple flat directions}

We now turn to the system with potential (\ref{pot5}). It admits two flat directions, characterized by the background values
\begin{equation}
\langle\phi_1\rangle = \langle\phi_2\rangle = \frac{F}{2 R}e^{i\Sigma} \;\;\;,\;\;\;
\langle\phi_3\rangle = \langle\phi_4\rangle = \frac{G}{2 R}e^{i\tilde\Sigma} \;\;\;,\;\;\;
\langle A_\mu \rangle = 0
\label{backval2}
\end{equation}
As in the single flat direction case, we work in the phenomenologically relevant limit of
\begin{equation}
\left\{ m \, R ,\, {\tilde m} \, R ,\, \Sigma' ,\, {\tilde \Sigma}' ,\, \frac{F'}{\sqrt{F^2+G^2}} ,\, 
\frac{G'}{\sqrt{F^2+G^2}} \right\} \ll \left\{ F ,\, G \right\}
\label{seriesapproximation}
\end{equation}
and we denote by $\epsilon$ the ratio between any of the terms on the left hand side of (\ref{seriesapproximation}) and any of the terms on the right hand side in this expression.

The fluctuations $\delta \phi_i$ of the scalar fields encode $8$ real degrees of freedom. One of them is 
set to zero in the unitary gauge; the other seven degrees of freedom can be written in the form 
\begin{eqnarray}
\delta \phi_1 + \delta \phi_2 = \frac{1}{R} \left( \delta_r + i \, \delta_i \right) \;\;\;,\;\;\;
\delta \phi_1 - \delta \phi_2 = \frac{{\rm e}^{i \Sigma}}{R} \left( \delta_H + i \, F \, a \right) \nonumber\\
\delta \phi_3 + \delta \phi_4 = \frac{1}{R} \left( {\tilde \delta}_r + i \, {\tilde \delta}_i \right) \;\;\;,\;\;\;
\delta \phi_3 - \delta \phi_4 = \frac{{\rm e}^{i {\tilde \Sigma}}}{R} \left( {\tilde \delta}_H - i \, G \, a \right) 
\label{deco2u1main}
\end{eqnarray}
We note that the same field $a$ enters in $\delta \phi_1 - \delta \phi_2$ and $\delta \phi_3 - \delta \phi_4$. The difference between these two entries is precisely the Goldstone boson, which has been set to zero in the unitary gauge. 

The modes $\delta_r$ and $\delta_i$ are the real and imaginary excitations of the first direction. They are decoupled from the other modes at the linearized level, and their mass matrix is identical 
to (\ref{massmatflat}). Analogously, the modes ${\tilde \delta}_r$ and ${\tilde \delta}_i$ are the real and imaginary excitations of the second direction. Their action is formally identical to that of $\delta_r$ and 
$\delta_i$, upon the substitution of the background quantities and the model parameters related to the first direction with the corresponding quantities of the second direction.

As in the previous case, we decompose the vector field into transverse and longitudinal components. The two transverse components are decoupled at the linearized level, and their mass is
\begin{equation}
m_{\rm gauge}^2 = \frac{e^2 \, \left( F^2 + G^2 \right)}{4 \, R^2}
\end{equation}

In summary, the above modes behave identically to those corresponding to the single flat direction; therefore, none of them are produced nonperturbatively. The action for the remaining modes, however, has no counterpart in the single flat direction case, and, as we now show, leads to nonperturbative production. Once the non-dynamical mode $A_0$ is integrated out, we are left with four physical modes: the longitudinal vector polarization $L$, and the three fluctuations encoded in the second line of  (\ref{deco2u1main}). 

The two fields $a$ and $L$ are not canonical. In Appendix \ref{app:twoU1} we give the two linear combinations $L_1 ,\, L_2$ which are canonical in terms of $a$ and $L$ so that in matrix form, the action for the system is
\begin{equation}
 S_{\rm coupled} = \frac{1}{2}\int d\eta \,d^3k \left(\Delta^{\prime\dagger} \Delta' + \Delta^{\prime \dagger} K \Delta - \Delta^\dagger K \Delta^\prime -\Delta^\dagger \Omega^2 \Delta\right)\,,
\label{actfour-2fd}
\end{equation}
where we defined $\Delta \equiv \left(\delta_H \;,\; \tilde{\delta}_H \;,\; L_1 \;,\; L_2 \right)$, and where the matrices $\Omega^2$ and $K$ are real and, respectively, symmetric and anti-symmetric. The exact expressions for these matrices are rather involved. In Appendix \ref{app:twoU1} we present them 
as an expansion series in $\epsilon$ (defined in eq.~(\ref{seriesapproximation})).  

To compute the nonperturbative production of these modes, we need to find the eigenmasses of the physical modes, as well as the matrix $\Gamma \,$, defined in eq.~(\ref{defineIJGamma}). For the eigenmasses, we find
\begin{eqnarray}
m_1^2 &=& \frac{e^2 \, \left( F^2 + G^2 \right)}{4 \, R^2} + \left[ \frac{\left( F^2 \, m^2 + G^2 \, {\tilde m}^2 \right) }{F^2 + G^2} + \frac{3 \left( F^2 \Sigma' + G^2 {\tilde \Sigma}' \right)^2}{R^2 \left( F^2 + G^2 \right)^2} \right] + \mathcal{O} \left( \epsilon^4 \right) \nonumber\\
m_2^2 &=& \left[ \frac{\left( F^2 \, {\tilde m}^2 + G^2 \, m^2 \right) }{F^2 + G^2} + \frac{3 \left( F \, G' - G \, F' \right)^2}{R^2 \left( F^2 + G^2 \right)^2} + \frac{3 \, F^2 \, G^2 \left( \Sigma' - {\tilde \Sigma}' \right)^2}{R^2 \left( F^2 + G^2 \right)^2} \right] + \mathcal{O} \left( \epsilon^4 \right) \nonumber\\
m_3^2 &=& \left[ \frac{\left( F^2 \, {\tilde m}^2 + G^2 \, m^2 \right) }{F^2 + G^2} \right] + 
\mathcal{O} \left( \epsilon^4 \right) \nonumber\\
m_4^2 &=& \frac{e^2 \, \left( F^2 + G^2 \right)}{4 \, R^2} + \mathcal{O} \left( \epsilon^4 \right)
\label{masseigenvalues}
\end{eqnarray}
where terms outside of the parenthesis, and inside square parenthesis are, respectively, of zeroth and second order in $\epsilon \,$. We see that the system has two heavy and two light eigenstates. The fourth eigenmass coincides with that of the transverse vector modes (up to the accuracy of the present computation) and therefore we refer to the corresponding eigenstate as the physical longitudinal vector mode. The first eigenstate is instead the Higgs field of the model. As in the single flat direction case, these two fields have identical mass at leading order. 

For a single flat direction, the Higgs and the longitudinal vector mode were coupled only amongst each other in the action (\ref{az1U1LH0s}). Now, they are coupled to the additional light modes, which are absent in the single flat direction case. It is easy to verify that the eigenfrequencies of the light modes vary nonadiabatically with time. Namely, starting from the four eigenmasses (\ref{masseigenvalues}), we compute the comoving frequencies $\omega_i = \sqrt{R^2 m_i^2 + k^2} \,$, and we compute the adiabatic conditions $\omega_i' / \omega_i^2 \,$ for momenta $k$ comparable with the flat direction masses (hence, of order $\epsilon$ in our notation). For the two heavy eigenstates, we find
\begin{equation}
\frac{\omega_1'}{\omega_1^2} \;,\; \frac{\omega_4'}{\omega_4^2} = \mathcal{O} \left( \epsilon \right)
\label{adiaheavy}
\end{equation}
This agrees with the general expression (\ref{resultpot1}) valid for the heavy fields. However, for the light modes we find
\begin{equation}
\frac{\omega_2'}{\omega_2^2} \;,\; \frac{\omega_3'}{\omega_3^2} = \mathcal{O} \left( 1 \right)
\label{nonadialight}
\end{equation}
Hence, we see that the time variation of these eigenfrequencies is not suppressed in the limit of $m \ll \vert \Phi \vert$ ! As the numerical results presented in the next section show, this leads to the strong nonperturbative production of these modes.

A second source of nonadiabaticity comes from the mixing in field space encoded in the (anti-symmetric) matrix $\Gamma$. We find, 
\begin{eqnarray}
&& \Gamma_{13} = - \left\{ \frac{\sqrt{\left( F \, G' - G \, F' \right)^2 + F^2 \, G^2 \left( \Sigma' - {\tilde \Sigma}' \right)^2}}{F^2 + G^2} \right\} + \mathcal{O} \left( \epsilon^2 \right) \nonumber\\
&& \Gamma_{23} = \left\{ \frac{R^2 \left( m^2 - {\tilde m}^2 \right) \, F^2 \, G^2 \, \left( \Sigma' - {\tilde \Sigma}' \right) }{\left( F \, G' - G \, F' \right)^2 + F^2 \, G^2 \left( \Sigma' - {\tilde \Sigma}' \right)^2} - \frac{F^2 \, \Sigma' + G^2 \, {\tilde \Sigma}'}{F^2 + G^2} \right\} + \mathcal{O} \left( \epsilon^2 \right) \nonumber\\
&& \Gamma_{12} ,\, \Gamma_{14} ,\, \Gamma_{24} ,\, \Gamma_{34} = \mathcal{O} \left( \epsilon^2 \right) 
\label{gammamatrix}
\end{eqnarray}
The terms within curly brackets are first order in $\epsilon \,$. As a consequence,  two of the off-diagonal nonadiabaticity coefficients, defined in 
eq.~(\ref{adiabaticmatrix}), are large
\begin{equation}
A_{13} ,\, A_{23} = \mathcal{O} \left( 1 \right)
\label{a13a23}
\end{equation}
while the remaining ones are suppressed by an additional factor of $\epsilon$. This also leads to strong particle production (in particular, due to the $A_{13}$ term, also the heavy Higgs mode is produced).

We conclude this Section by noting that the physical longitudinal vector mode (more appropriately, the fourth eigenstate) has a negligible coupling with the other three modes of the system. This mode does not experience nonadiabatic production.



\section{Numerical Results}
\label{sec:results}

We perform the numerical computation of particle production in the two flat direction case, with potential
(\ref{pot5}). We focus on the system of fields $\left\{ \delta_H ,\, {\tilde \delta}_H ,\, L_1 ,\, L_2 \right\}$ (as defined in the previous Section) for which nonperturbative production can be relevant. The background evolves as described in eqs. ~(\ref{eombck}), with two additional equations for the amplitude and phase of the flat directions (identical in form as the first two of (\ref{eombck})), and with the potential of both flat directions appearing in the equation for the scale factor. Particle production is obtained from eqs. (\ref{bogequations}).

The timescale for the evolution of the flat directions is set by their mass, so that terms of order $\epsilon$ must be retained in the numerical equations solved. However, terms of higher order can be neglected. The coupling between the different modes is encoded in the matrix $\Gamma$, whose entries for the system we are studying are given in (\ref{gammamatrix}). We then see that, if we neglect terms higher order in $\epsilon$, the longitudinal vector mode decouples from the other three. The mass of this mode varies adiabatically with time. Therefore, we can disregard this decoupled field in the computation, and we focus on the system of the three remaining modes.

We should note that the potential (\ref{pot5}) is not bounded from below, due to the presence of the quartic terms proportional to $\lambda$ and ${\tilde \lambda}$. Such terms dominate over the quadratic ones for large field values. In a complete model, we expect that higher order terms will be also present, and stabilize the potential. In our computation, we simply choose the parameters such that the quartic terms are subdominant (their presence is however crucial to generate the rotation of the flat directions in their complex planes). As we already discussed, the quartic terms 
can arise from one loop contributions to the effective potential and are assumed to have magnitude $\lambda\sim \frac{h^2 m^2}{\vert\Phi_0\vert^2}$ (and, analogously, for the second direction) in analogy to the baryon number violating operators in the Affleck-Dine mechanism \cite{ad}, where $h^2 \sim 0.1$ is some coupling constant. For definiteness, we set
\begin{equation}
\lambda = \frac{m^2}{10 \, \vert \Phi_0\vert^2} \quad\,,\quad \tilde{\lambda} = \frac{\tilde{m}^2}{10 \, \vert \tilde{\Phi}_0\vert^2}\,.
\end{equation}
This ensures that the quartic terms are initially subdominant, and that the two flat directions evolve towards the origin.

We introduce the following dimensionless quantities (each denoted with an asterisk), 
which are used in the numerical evolution:
\begin{eqnarray}
&& \eta_* \equiv e \vert \Phi_0 \vert \eta \nonumber\\
&& m_* \equiv \frac{m}{e \vert \Phi_0 \vert} \;\;\;,\;\;\;
{\tilde m}_* \equiv \frac{{\tilde m}}{e \vert \Phi_0 \vert} \;\;\;,\;\;\;
k_* \equiv \frac{k}{e \vert \Phi_0 \vert} 
\;\;\;,\;\;\; \lambda_* \equiv \frac{\lambda}{e^2} \;\;\;,\;\;\; {\tilde \lambda}_* \equiv \frac{\tilde \lambda}{e^2}
\nonumber\\
&& F_* \equiv \frac{F}{\vert \Phi_0 \vert} \;\;\;,\;\;\; G_* \equiv \frac{G}{\vert \Phi_0 \vert} \;\;\;,\;\;\;
\rho_{\psi*} \equiv \frac{\rho_\psi}{e^2 \, \vert \Phi_0 \vert^2 M_p^2}
\label{program}
\end{eqnarray}
The background equations, and the equations for particle production in dimensionless quantities are given, respectively, in eqs. (\ref{bckprog}) and (\ref{bogprog}). One can see that the value of $e$ factors out in these equations, and we only need to specify the value of $\vert \Phi_0 \vert / M_p$ in the equation for the scale factor. 

We are free to set the initial value for the scale factor $R_0 = 1 \,$, which implies
\begin{equation}
F_{*0} = 2 \;,\;\; G_{*0} = 2 \, 
\frac{\vert {\tilde \Phi}_0 \vert}{\vert \Phi_0 \vert}
\label{initialprg}
\end{equation}
As we discussed in the introduction, flat directions can influence the thermal history of the universe only if they have a large initial amplitude. Therefore, we choose $\vert \Phi_0 \vert / M_p = 10^{-2} \,$ in our numerical analysis. We also start with the flat directions at rest, an initial condition ascribed  to Hubble friction.~\footnote{As clear from the parametrization (\ref{backval2}), this implies $\left( F / R \right)' = \Sigma' = \left( G / R \right)' = {\tilde \Sigma}' = 0 \,$. The initial value for $R'$ is obtained from the Friedmann equation.} At early times, the energy density of the universe is dominated by the inflaton contribution $\rho_{\psi*}$. As such, we start from the initial value $\rho_{\psi*} = 1 \,$ (see for example, the last expression of (\ref{bckprog})). As discussed in \cite{op}, for a gravitational inflaton decay, the inflaton has yet to decay when the flat directions start evolving. At this stage, the inflaton is oscillating around the minimum of its potential, so that its energy density decreases as that of matter; therefore, we set $\rho_{\psi*} = 1 / R^3$ in our numerical computations.

\begin{figure}[th]
\begin{center}
(a) \epsfig{file=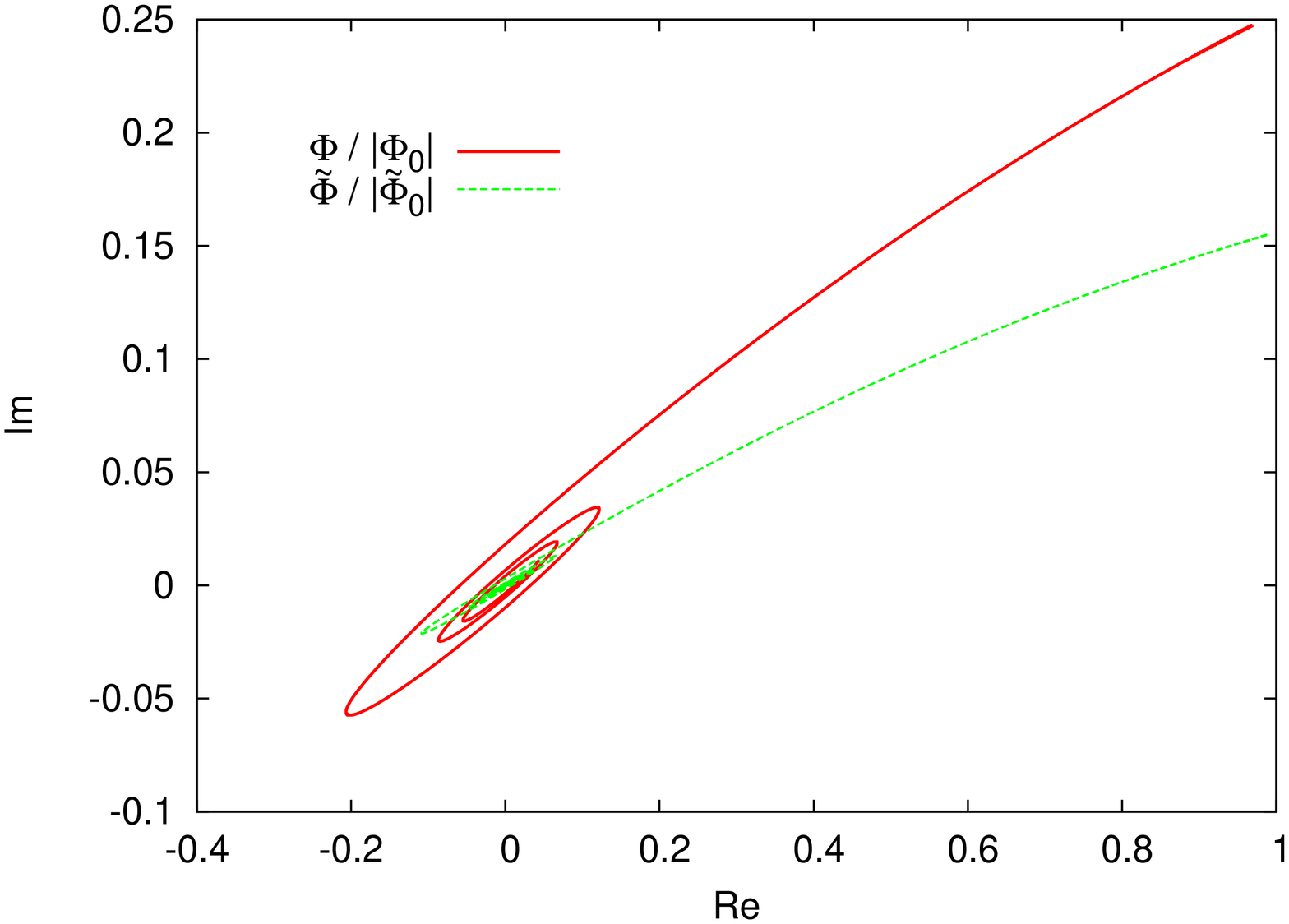,width=5.8cm,angle=270}
(b) \epsfig{file=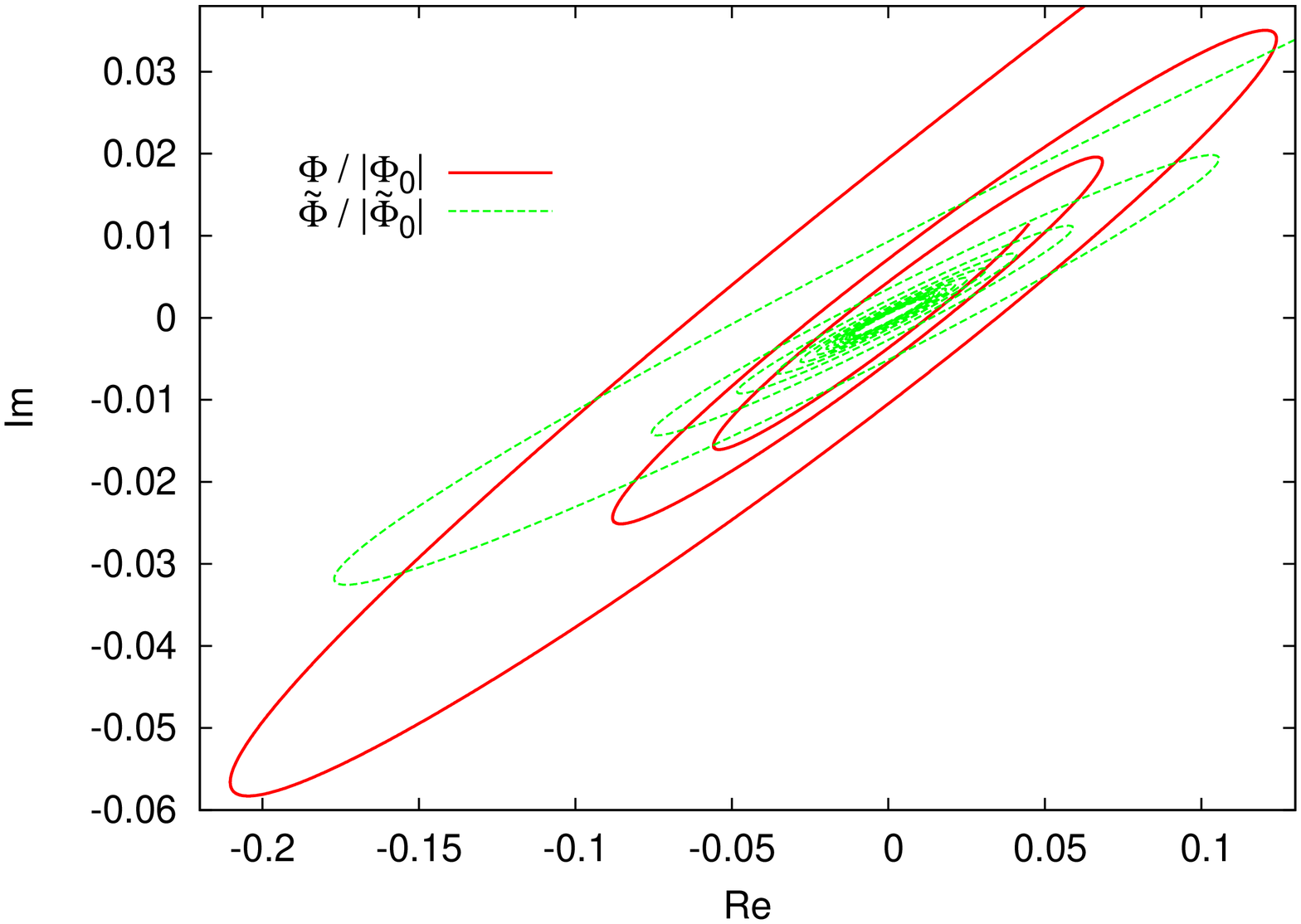,width=5.8cm,angle=270}
\caption{\label{fig:background} The background evolution of the flat directions 
for the choice of masses $m_* = 
10^{-6} ,\; \tilde{m}/m = 3.72$ and initial phases $\Sigma_0 = 0.25$ , 
$\tilde{\Sigma}_0 = 0.156$. The initial values of the vevs are chosen such 
that $|\tilde\Phi_0| / |{\Phi}_0| = 15$ and $|{\Phi}_0| = 10^{-2} M_p$.
We show in the real and imaginary parts of the flat directions for the 
first three rotations of $|\Phi_0|$ (solid red) and $\vert\tilde\Phi\vert$ (solid green);
the right panel shows the same evolution as in the left panel, zoomed in on smaller values
of the fields. }
\end{center}
\end{figure}

For illustrative purposes, we show in figure \ref{fig:background} the evolution of the flat directions for some specific choice of parameters. We fix the soft masses to $m_*=10^{-6}$, $\tilde{m}/m = 3.72$ and numerically evolve the background equations (\ref{bckprog}) with initial values $F_{*0}=2$, $G_{*0}=30$ and $\Sigma_0 = 0.25$, $\tilde{\Sigma}_0 = 0.156$. In the left panel we show the evolution of the real and imaginary parts of the flat direction vevs, normalized to their own initial values. In the right panel, we zoom in on the values closer to the origin, to better visualize the spiral motion of the two fields.

We set the initial conditions for the Bogoliubov coefficients as $\alpha_0 = \mathbf{1}$, $\beta_0 = 0$, 
corresponding to an initial absence of particles. At the start of the background evolution, the masses of the light degrees of freedom are dominated by the tachyonic contribution $R'' / R^3$. This contribution rapidly becomes subdominant, and can be neglected when the flat directions start evolving. Our formalism for the production is valid only as long as the eigenfrequencies are positive. For this reason, we start the evolution of (\ref{bogequations}) only once the eigenmasses have all become positive (we stress that the flat directions are still frozen at this stage).This amounts to neglecting the small gravitational particle production during the initial stages.

From the numerical evolution, we obtain the occupation number densities $n_i$ of produced quanta, 
eq.~(\ref{occupationnumber}). We can then compute the ratio between the energy density in these quanta, and that in the flat directions; in terms of the dimensionless quantities defined in (\ref{program}), the ratio is
\begin{eqnarray}
r_{\rm prod} \equiv \frac{\rho_{\rm prod}}{\rho_{\rm flat}} &=& \left(16\,\pi\,e^2\right) \left(\sum_{i=1} ^3 \int dk_*\,k_*^2 \omega_{i*} (k_*) n_i (k_*)\right)\nonumber\\
&&\times \left\{ F_*^2 \left[ \left(\frac{F_*'}{F_*} -\frac{R'}{R} \right)^2 +\Sigma'^2 \right] +R^2 F_*^2 \left[ m_*^2 + \frac{\lambda_*}{4} \, \frac{F_*^2}{R^2} \cos(4 \,\Sigma) \right] \right.\nonumber\\
&&  \quad\quad\left. +\;G_*^2 \left[ \left(\frac{G_*'}{G_*} -\frac{R'}{R} \right)^2 +\tilde{\Sigma}'^2 \right] +R^2 G_*^2 \left[ \tilde{m_*} ^2 + \frac{\tilde{\lambda_*}}{4} \, \frac{G_*^2}{R^2} \cos(4 \,\tilde{\Sigma}) \right] \right\}^{-1}\,.
\label{energydensityratio}
\end{eqnarray}
where the dimensionless eigenfrequencies are defined as $\omega_{i*} \equiv \omega_i / \left( e \vert \Phi_0 \vert \right)\,$. If $r_{\rm prod}$ becomes equal to one, we say that the flat directions have decayed. The present analysis is actually invalid at this point, since we are ignoring the backreaction of the produced quanta on the evolution of the flat directions (in particular, we do not account for the decrease in the amplitudes of the flat directions due to particle production). However, setting $r_{\rm prod} = 1$ allows us to understand for which choice of parameters particle production is significant, and to estimate the decay time of the flat directions. It is useful to use as a measure of time the number of rotations of one of the two flat directions; for definiteness, we choose: $N \equiv \left( \Sigma - \Sigma_0 \right) / 2 \pi \,$. We denote by $N_{\rm decay}$ the value of $N$ at which the ratio $r_{\rm prod}$ equals to one. 

As we have mentioned, there are two different scales in this problem, set by the amplitude and the mass of the flat directions (the former controls the heavier eigenmass $m_1$ in (\ref{masseigenvalues}), while the latter the two light masses $m_{2,3} \,$). The timescale for particle production is set by the evolution of the flat directions, which in turn is governed by their mass; however, the intermediate matrices $\alpha$ and $\beta$ evolve on a much quicker timescale, set by the largest eigenmass $m_1 \,$. Both scales need to be under control in the numerical simulations. The ratio between these two scales is the quantity 
$\epsilon$ defined after eqs. (\ref{seriesapproximation}). In realistic cases, these two scales differ by many orders of magnitude (GUT or Planck, vs. TeV scale). Simulations with such small values of $\epsilon$ cannot be performed. Fortunately, as we discuss in appendix \ref{app:scaling}, the occupation numbers of the two fields exhibit an approximate but accurate scaling with the masses and the amplitudes of the two directions. If we rescale both masses and both amplitudes by a constant factor
\begin{eqnarray}
\{\Phi_0,\tilde\Phi_0\}&\rightarrow&\{\gamma\Phi_0,\gamma\tilde\Phi_0\} \,,\nonumber\\
\{m,\tilde m\}&\rightarrow& \{ \mu m,\mu \tilde m \} 
\label{rescalings}
\end{eqnarray}
then the occupation numbers scale as
\begin{equation}
\left\{ n_1 \;,\; n_2 \;,\; n_3 \right\} \rightarrow \left\{ \frac{\mu}{\gamma} \, n_1 \;,\; n_2 \;,\; n_3 \right\}
\end{equation}

\begin{figure}[th]
\begin{center}
\epsfig{file=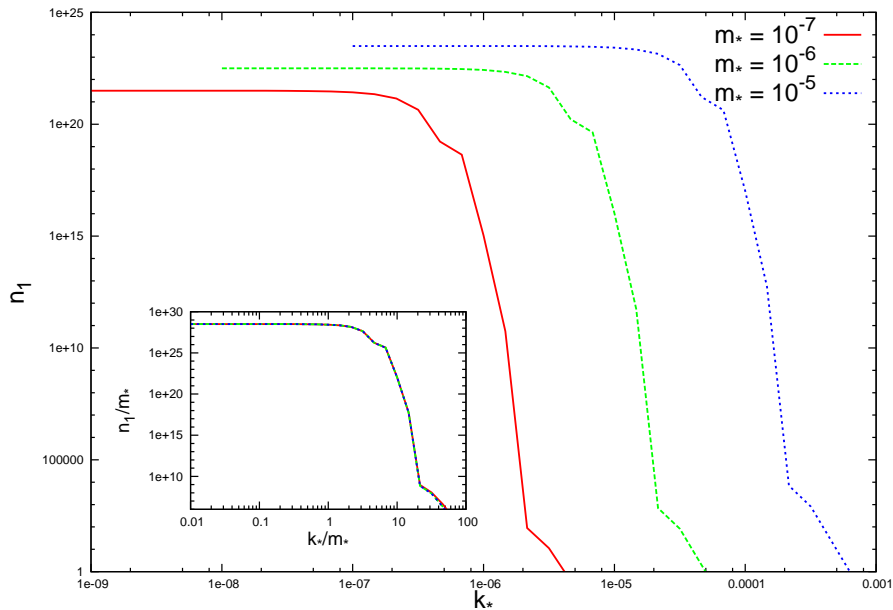,width=8.3cm,angle=270}
\caption{\label{fig:scaling} 
Occupation number for the first (heavy) eigenstate after three rotations of the first direction, for a specific set of parameters: $\vert \Phi_0 \vert = 10^{-3} \, M_p$, $\vert {\tilde \Phi}_0 \vert / \vert \Phi_0 \vert = 4$, ${\tilde m} / m = 3.72$, $\Sigma_0 = 0.25$, ${\tilde \Sigma}_0 = 0.156$. The larger figure shows the occupation number as a function of the momentum (rescaled as in eq.~(\ref{program})) for three specific choices of the (rescaled) flat direction mass $m_*$. The insert shows the same result, but plotting $n_1/ m_*$ vs. $k_* / m_*$. In terms of these variables, the three curves overlap. Since $m_* \propto \epsilon$, this confirms the scaling (\ref{scalingn}) for the occupation number, and the fact that the momenta of the quanta produced also scale as $\epsilon \,$.
}
\end{center}
\end{figure}

Since $\epsilon \rightarrow \left( \mu / \gamma \right) \epsilon$ under (\ref{rescalings}), this is equivalent to saying that
\begin{equation}
n_1 \propto \epsilon \;\;\;,\;\;\; n_{2,3} \propto \epsilon^0
\label{scalingn}
\end{equation}
The numerical solutions obtained in the range $10^{-7} \lta \epsilon \lta 10^{-5}$ confirm this scaling behavior, as we show for a specific set of parameters in figure \ref{fig:scaling}. We show the spectrum of the first (heavy) eigenstate for three distinct cases, which differ from each other only by the ratio between the masses and the amplitudes of the two directions (hence, only by the value of $\epsilon$). The three spectra coincide once $n_1$ is rescaled according to eq.~(\ref{scalingn}), and once the momentum is rescaled as $k / \epsilon \,$. Although we do not show it here, we also verified that the occupation numbers for the light eigenstates do not change with $\epsilon$. The system of equations for particle production is too involved to prove this scaling analytically. However, in appendix \ref{app:scaling}, we see that the scaling ``emerges'' from the equations under a few assumptions ($\vert \epsilon \, \alpha_{1j}' \vert \ll \vert {\bar \omega}_1 \, \alpha_{1j} \vert$ and $\vert \epsilon \, \beta_{1j}' \vert \ll \vert {\bar \omega}_1 \, \beta_{1j} \vert$, for $j=1,2,3$) which are verified numerically.

Perhaps the most convincing argument in favor of the scaling (\ref{scalingn}) is the fact that the
nonadiabaticity parameters for the light states, $\omega_2' / \omega_2^2$ and $\omega_3' / \omega_3^2$, are not suppressed in the limit of small $\epsilon \,$, cf. eqs.~(\ref{nonadialight}). 
The analogous parameter for the heavy state is suppressed. However, the production of this state takes place through the coupling with the two lighter ones (cf. eq.~(\ref{a13a23}). Since $\omega_1 \propto \epsilon^0 ,\, \omega_{2,3} \propto \epsilon \,$, equipartition of energy then suggests that $n_1$ is suppressed by $\epsilon$ with respect to $n_{2,3} \,$.

Due to these scaling properties, the ratio between the energy density of the produced quanta and that of the background flat directions scales as (see appendix \ref{app:scaling})
\begin{equation}
r_{\rm prod} \rightarrow 
\frac{\mu^2}{\gamma^2}\,r_{\rm prod}
\end{equation}
This implies that $r_{\rm prod}$ is of the form
\begin{equation}
r_{\rm prod} \simeq \frac{\tilde{m} \,  m }{ |\tilde{\Phi}_0| \, |\Phi_0|} \times f \left( \frac{\tilde m}{m} ,\, 
\frac{\vert {\tilde \Phi}_0 \vert}{\vert \Phi_0 \vert} ,\, N \right)
\label{functionf}
\end{equation}
The multiplying function $f$ can be computed numerically for cases in which the amplitudes and the masses of the flat directions differ only of a few orders of magnitude (the approximate scaling holds only as long as $\epsilon \ll 1$). The scaling (\ref{functionf}) then gives the production for realistic cases ($\epsilon$ as small as $10^{-15}$) for which numerical evolution is not feasible.

The function $f$ obtained numerically exhibits a strong dependence on the parameters. The rate of particle production grows exponentially within a range of parameters, while it is negligible otherwise. Therefore, eq. (\ref{functionf}) can be cast in the form
\begin{equation}
r_{\rm prod}  \simeq C \,
\frac{\tilde{m} \,  m }{ |\tilde{\Phi}_0| \, |\Phi_0|} \, 10^{\,\sigma 
\, N} \,,
\label{expgrowth}
\end{equation}
where $C$ and $\sigma$ are two time--independent quantities that are functions only of the ratios $\tilde{m}/m$ and $| \tilde{\Phi}_0|/| \Phi_0|$. The growth rate, $\sigma$, rapidly approaches zero outside the range for which particle production takes place.

\begin{figure}[th]
\begin{center}
\epsfig{file=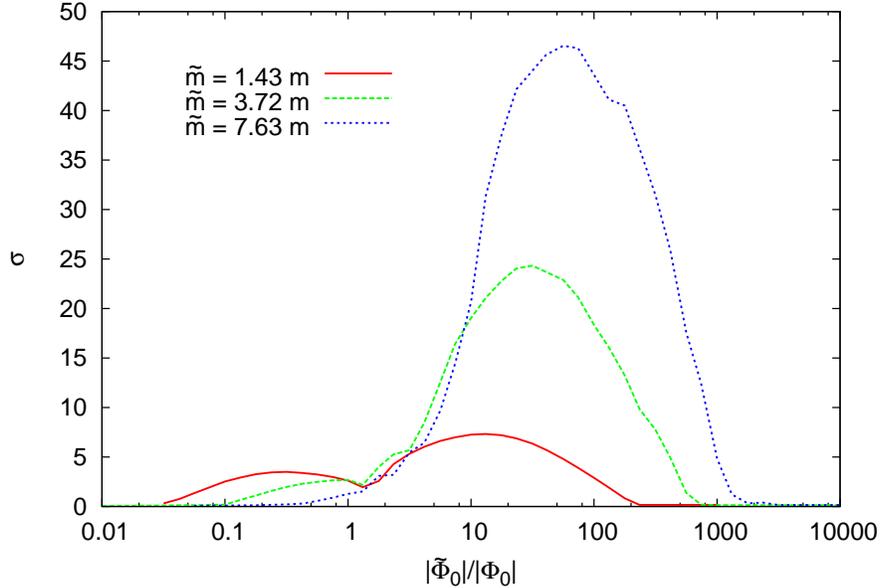,width=8.3cm,angle=270}
\caption{\label{fig:growthexp} The dependence of the growth exponent $\sigma$ 
defined in (\ref{expgrowth}) on the initial ratio of the vevs $| 
\tilde{\Phi}_0|/| \Phi_0|$, for three different mass ratios.}
\end{center}
\end{figure}

In figure \ref{fig:growthexp}, we show the numerical values for $\sigma$ obtained from the numerical simulations for three given mass ratios, and for a range of ratios between the two amplitudes.

\begin{figure}[th]
\begin{center}
\epsfig{file=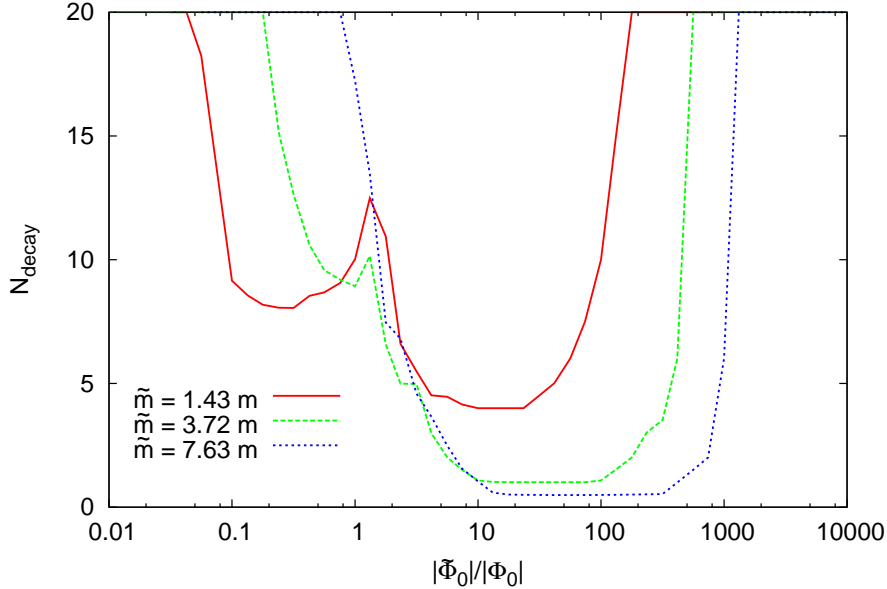,width=8.3cm,angle=270}
\caption{\label{fig:production} Number of rotations of the first vev where the 
production criterion $r_{\rm prod} = 1$ is 
satisfied and its dependence on the initial ratio of the vevs $| 
\tilde{\Phi}_0|/| \Phi_0|$, for three different mass ratios.  The numerical 
analysis was made for the first twenty rotations of the first flat direction 
and the results were rescaled to $m/e= 10^4\,{\rm GeV}$ and $ \sqrt{| \tilde{\Phi}_0| 
\, | \Phi_0|} = 10^{-2} M_p $.}
\end{center}
\end{figure}

As is clear from (\ref{expgrowth}), $N_{\rm decay}$ is inversely proportional to $\sigma \,$,
\begin{equation}
N_{\rm decay} \simeq \frac{1}{\sigma} \, {\rm log}_{10} \left(\frac{| 
\tilde{\Phi}_0| \, | \Phi_0|}{ C\,\tilde{m}\,m} 
\right)\,.
\label{ndecay}
\end{equation}
We show this quantity in figure \ref{fig:production}, for the same ratios of masses and amplitudes used in figure \ref{fig:growthexp}. The overall mass scale is set to $m/e = 10^4 \, {\rm GeV}$ (the mass of the first direction). The overall scale of the amplitudes is fixed by rescaling the results to $\left( \vert \Phi_0 \vert \, \vert {\tilde \Phi}_0 \vert \right)^{1/2} = 10^{-2} \, M_p \,$.

\begin{figure}[th]
\begin{center}
\epsfig{file=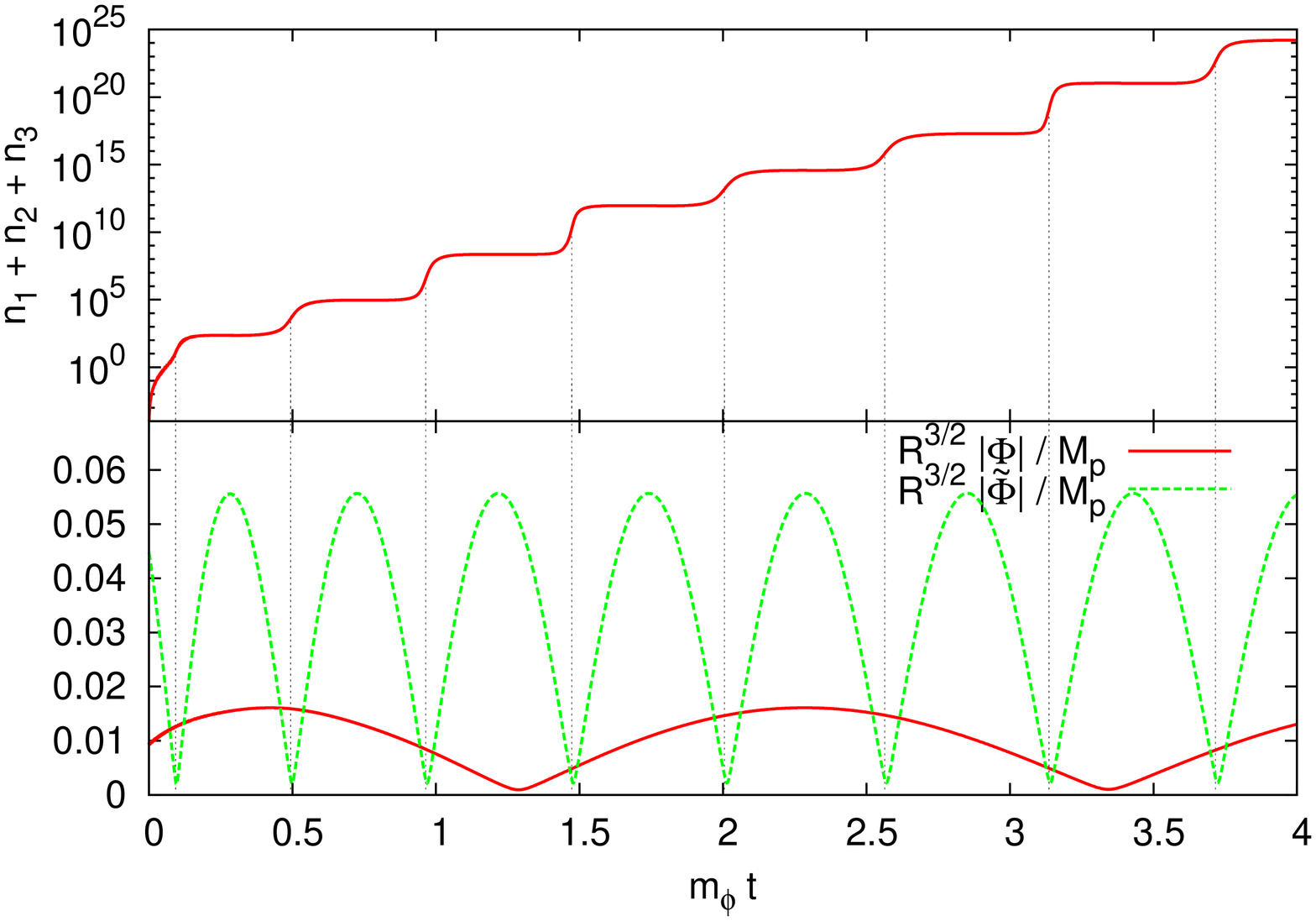,width=8.3cm,angle=270}
\caption{\label{fig:correlation} Upper panel: Number density of the produced quanta for a given momentum $k_* = 10^{-7} \,$. Lower panel: amplitudes of the two flat direction. We note that particle production occurs whenever the two amplitudes are comparable, as discussed in the main text. The parameters chosen for this evolution are $m_* = 10^{-6} ,\, F_* = 2 ,\, {\tilde m} / m = 3.72 ,\, \vert {\tilde \Phi}_0 \vert / \vert \Phi_0 \vert  = 15$}
\end{center}
\end{figure}

We see that the production of particles is extremely fast, provided the initial amplitudes of the two flat directions are not too different. Not surprisingly, a flat direction with a significantly smaller amplitude can be neglected, so that one is effectively back to the single flat direction case, for which we know that particle production is absent. This can be seen more clearly in figure \ref{fig:correlation} where we show the sum of the number densities of the produced quanta (for a given value of the momentum) in the upper panel, and the evolution of the amplitudes of the flat directions in the lower panel. We note that particle production takes place in a stepwise manner, whenever the amplitudes of the two directions are comparable to each other. The overall growth is exponential with time, as parametrized in equation (\ref{expgrowth}), once we average over complete rotations (this is also how the quantity $\sigma$, plotted in figure \ref{fig:growthexp} has been obtained).

The fact that particle production takes place whenever the two directions have comparable amplitude does not mean that the two amplitudes need to be equal (or nearly equal) initially. This is due to the fact that the orbits of the two flat directions are strongly elliptical (cf. figure \ref{fig:background}), so that each amplitude varies significantly during its rotation. As a result, we see in figure \ref{fig:production} that production takes place for a significantly wide range (about four orders of magnitudes) of the ratio of the initial amplitudes. We actually see that the central point of this interval moves to larger values of $\vert {\tilde \Phi}_0 \vert / \vert \Phi_0 \vert$ as ${\tilde m} / m$ increases. To understand this, imagine that the second direction is heavier than the first one. Then, there is an interval of time during which the first direction is frozen, while the second one evolves (these are the times for which $m < H \left( t \right) < {\tilde m}$). During this interval, the amplitude of the second direction (more appropriately, the maxima and the minima of the amplitude) decreases, due to Hubble friction, as 
$R^{-3/2}$, while the amplitude of the first direction remains constant. Therefore, to have maximal overlapping between the two amplitudes when both directions are evolving, the second direction should start with a higher amplitude than the first one. The value of $\vert {\tilde \Phi}_0 \vert / \vert \Phi_0 \vert$  leading to maximal overlapping (and maximal production) increases as ${\tilde m} / m$ increases, as figure \ref{fig:production} clearly shows.

\begin{figure}[th]
\begin{center}
\epsfig{file=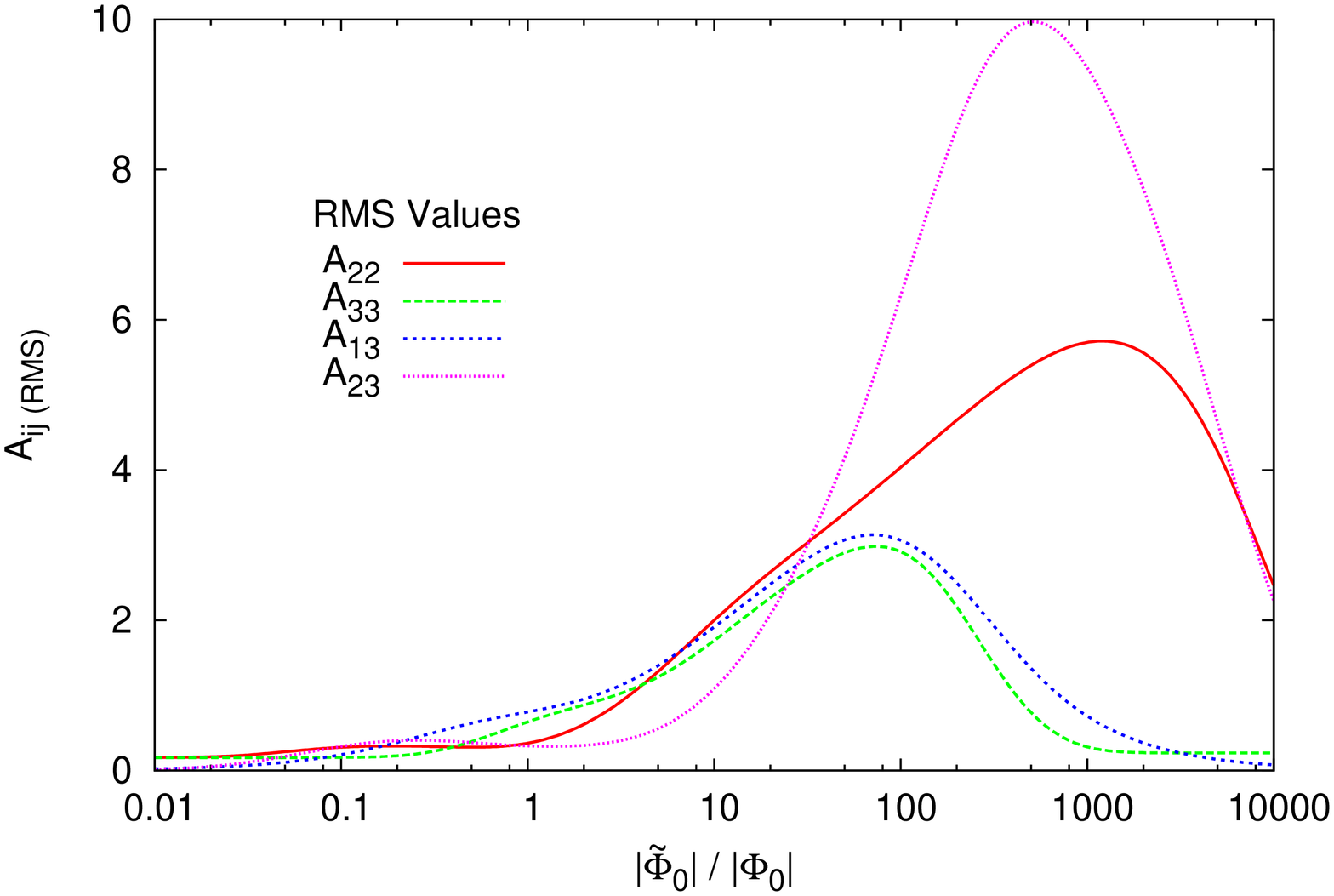,width=8.3cm,angle=270}
\caption{\label{fig:adiabaticmatrix} The root mean square (over physical time) 
of the four leading adiabatic matrix elements $A_{22},A_{33},A_{13}$ and 
$A_{23}$ of the modes with momentum $k=0$, for five rotations of the flat 
direction $\Phi$. While the RMS values are not directly correlated with the 
growth exponent $\sigma$, they indicate the parameter space where particle 
production may be possible.}
\end{center}
\end{figure}

Finally, we observe that the large production of particles can be traced to the adiabatic parameters. The leading order adiabatic matrix elements $A_{22},A_{33},A_{13}$ and $A_{23}$ are all of order one in the series approximation~(\ref{seriesapproximation}) and generically are greater than one during much of the evolution. We show the root mean square (RMS) of the leading order adiabatic matrix elements in figure~\ref{fig:adiabaticmatrix}, for the mass ratio $\tilde{m}=3.72\,m$.  As explained in Sections~\ref{sec:models}~and~\ref{sec:formalism}, a necessary (but not sufficient) condition for the production of quanta  is realized when any element of the adiabatic matrix is greater than one.  By comparing figure~\ref{fig:production} with figure~\ref{fig:adiabaticmatrix}, we see that the results support this assertion. 

\section{Discussion and Conclusions}
\label{sec:discussion}

In this paper, we have computed the nonperturbative decay of systems of multiple flat directions. Our results agree with the conclusions found in \cite{op}. However, we improve over that study by including gauge fields in the computation. We worked out in detail the production of particles for systems with one or two U(1) flat directions. Specifically, we studied the linearized theory of fluctuations around such backgrounds, and we computed the production of particles due to the time dependency of the physical eigenstates and eigenmasses of the systems. We also showed that this computation can be readily applied to
SU(N) directions, as long as the gauge fields do not develop background expectation values.

Technical difficulties associated with this study revolved around the fact that one needs to find the spectrum of the fluctuations around a time dependent background. Firstly, one finds that, even in the unitary gauge, the longitudinal component of the gauge boson remains coupled to the ``matter'' fluctuations (the Higgs and light fields) at the linearized level. This coupling was not included in the analogous computations of \cite{basboll1,basboll2}, which adopt a gauge choice equivalent to that chosen here. However, this coupling needed to be included for consistency, and forced us to perform some field redefinitions (see Appendix \ref{app:twoU1} for details). 

Secondly, the quadratic action for the perturbations contains nonstandard quadratic kinetic terms, and ``mixed kinetic terms'' of the type $K_{ij} \delta \phi_i' \, \delta \phi_j' + X_{ij} \delta \phi_i' \, \delta \phi_j \,$, where the matrixes $K$ and $X$ depend on the background and are therefore time dependent. We provide a general formalism to diagonalize (and eventually quantize) such system, which is a necessary step to compute particle production. In particular, we show how particle production can be obtained without explicitly performing the rotation that eliminates the mixed term proportional to $X$. 

Thirdly, the resulting system of equations for particle production is too involved to be solved analytically. In the case of two flat directions, the relevant computation can be reduced to a complicated system of four coupled fields. To compute particle production, one needs to know the eigenvectors and the eigenmasses of this system. The analytic computation of these quantities is unpractical. On the other hand, the system possesses two mass scales, one given by the masses $m$ of the flat directions, and one by their amplitudes $\vert \Phi \vert \,$. While $m$ is expected to be of the order of the electroweak scale, $\vert \Phi \vert \,$ can be as large as the GUT or the Planck scale (indeed, very large vevs are necessary for the flat directions to play an important role in the thermal history of the universe \cite{op}). Any numerical evaluation of particle production needs to control both scales; this is also an untreatable problem, for realistic values of $\epsilon = m / \vert \Phi \vert \,$. Fortunately, the computation of the eigenmasses and eigenvectors can be performed analytically as an expansion series in $\epsilon \,$. We could then perform a numerical evaluation of the resulting formulae for intermediate values $10^{-7} \lta \epsilon \lta 10^{-5}$. The results of these evaluations showed that the particle production exhibits a clear scaling with $\epsilon$ (we also presented several analytic arguments in support of the observed scaling). The most important outcome is that the non adiabaticity in the evolution of the eigenmasses and the eigenvectors is not suppressed in the limit of $\epsilon \rightarrow 0 \,$, as was the case for the toy models considered in \cite{Allahverdi:1999je,Postma:2003gc,am1}. As a consequence, the flat directions can undergo very fast nonperturbative decay even in this limit.

The results of \cite{op} for the non-perturbative decay of flat directions have been discussed in some
recent papers \cite{am2,basboll1,basboll2,am3}. Ref. \cite{am2} agrees with \cite{op} that the effect is absent when only a single flat direction is excited; however, contrary to \cite{op}, it was argued there that the nonperturbative decay is unlikely in  the case of multiple flat directions. The main argument against this is that, according to \cite{am2}, (i) flat directions present hierarchical vevs, and (ii) there is no nonperturbative particle production whenever the vevs of the two directions differ by more than one order of magnitude (for more details, see their eq. (27), and the discussion immediately afterwards). The claim (i) is based on the fact that the higher order terms $\phi^n / M^{n-3}$ that we have discussed in the Introduction arise at different values of $n$ for different directions. As we already mentioned, the fact that one such term is allowed does not necessarily mean that it is present. On the contrary, we have already seen that, for the simplest inflaton model we are considering, all terms with $n < 11$ must be forbidden. Therefore, either such terms are present (in which case the vevs of the flat directions are too small), or are forbidden (in which case we cannot use this argument to see at which order any given direction is lifted). Also the claim (ii) is not substantiated by numerical computations, or by a clear proof. Our explicit computations show that this conclusion is actually incorrect, since we find particle production for a range of four orders of magnitude among the initial values of the two amplitudes.

In contrast, ref. \cite{basboll1} also finds that the effect of particle production can be present in the multiple flat directions case and agrees with \cite{op} . However, it is argued there that particle production can be present also for a
single flat direction. This conclusion is based on a specific example (their Section IV), where nonperturbative production takes place. Ref. \cite{basboll1} claims that there is one flat direction in this example. We agree that there is nonperturbative production in this model. However, this model contains two flat directions. This is clear from the fact that the background fields in this model have two physical (i.e., that cannot be removed by a gauge transformation) phases, their $\sigma$ and $\gamma$, that evolve independently with time (we count the number of flat directions, based on the number of independent physical background fields involved).

Ref. \cite{basboll2} studied whether realistic MSSM flat directions can undergo nonperturbative decay. It was concluded there that this effect does not take place for a single flat direction $LLe^c$ or $u^cd^cd^c$. It was also shown that nonperturbative decay can take place for the simultaneous presence of the directions $LLe^c$ and $QLd^c$, but not for the simultaneous presence of $LLe^c$ and $u^cd^cd^c$. These conclusions agree with those of \cite{op}. In particular, it is easy to understand why the nonperturbative decay does not take place when the $LLe^c$ and $u^cd^cd^c$ directions are excited. In the presence of these two directions, the $D-$term potential involves $28$ fields (the excitations of the fields which acquire nonvanishing background values, plus those related tho these fields by off-diagonal gauge generators). These two directions break all the SM gauge symmetries. As a consequence, there are $12$ goldstone bosons, and $12$ Higgses in the spectrum. These $24$ fields, plus the $4$ fields representing the real and imaginary excitations of the two flat directions, exhaust all the fields in the spectrum. Therefore, there are no additional light fields in which the heavy fields can rotate into. This is the reason why these two directions alone were not considered in \cite{op}; however, these two directions allow for the presence of a third one, of the $QLd^c$ type. In this case, there are sufficient fields in the spectrum to make nonperturbative decay possible.

Finally, ref. \cite{am3} claims that flat directions cannot undergo nonperturbative decay, due to some charge conservation. Ultimately, the conserved charge is the angular momentum of the background flat directions (in field space; for many directions this is related to the baryon or lepton number). We do not claim that the $D-$term interactions remove this charge. However, they can redistribute it from the homogeneous condensate ($k = 0$) to the quanta of the fields coupled to the condensate through the $D-$term potential. This effect is missed if one only considers the homogenous fields.

To conclude, we confirmed the results of \cite{op} that two or more flat directions possess one nonperturbative instability that can result in a much faster decay than what one would simply argue from their perturbative interactions. As it is typical for preheating studies, this instability is due to the nonadiabatic evolution of the eigenmasses and eigenvectors of the spectrum of the theory around the background flat directions. Once the production becomes significant, the nonlinear interactions of the produced quanta, and their backreaction on the homogeneous flat directions (besides the obvious decrease of the energy of the flat directions) will likely become important. This is a nontrivial issue, due to the complication of the MSSM potential \cite{op}, which deserves further study.

\begin{acknowledgments}
This work is partially supported by  the DOE grant DE-FG02-94ER-40823.
\end{acknowledgments}

\bigskip
\appendix


\section{One U(1) Flat Direction}
\label{app:oneU1}

In this appendix we study the flat direction composed of two scalar fields $\phi_1, \phi_2$ with the Lagrangian (\ref{lagrangian}) and the potential~(\ref{pot4}). Let us first comment on the simplest case of a U(1) symmetric theory with a single complex field $\chi$ (giving rise to the standard Higgs mechanism). Denoting, respectively, by $v$ and $\xi$ the amplitude and phase of this field, and taking the covariant derivative as 
in~(\ref{lagrangian}), with $q_\chi = 1\,$, one finds the following coupling between the scalar and the gauge fields
\begin{equation}
\vert D \chi \vert^2 \supset v^2 \left( \partial_\mu \xi - \frac{e}{2} \, A_\mu \right)^2
\end{equation}
It is then conventional to choose the unitary gauge in which the phase $\xi$ is set to zero. In this gauge, the linear coupling between the vector and the scalar is absent. If we then expand $v$ in a (nonvanishing) vacuum expectation value plus fluctuation, the spectrum of the theory consists of a Higgs field, and a massive gauge boson, which are decoupled from each other at the quadratic level.

Let us now turn to the case of our interest. We decompose the two fields entering in the potential~(\ref{pot4}) as 
\begin{equation} 
\phi_1 = \frac{f}{\sqrt{2} R} e^{i(\beta+\alpha)} \;\;\;,\;\;\;
\phi_2 = \frac{g}{\sqrt{2} R} e^{i(\beta-\alpha)} \nonumber
\end{equation}
where $f$, $g$, $\alpha$ and $\beta$ are real fields. The parametrization chosen for the phases is related to the fact that $\alpha$ is the only quantity changing under the U(1) transformation (since the two fields have opposite U(1) charge: $q_1 = 1 ,\, q_2 = -1$). We can completely fix the gauge freedom by going to the unitary gauge, in which $\alpha$ is set to zero. The coupling between the gauge and the scalar fields now reads
\begin{equation}
\vert D \phi_1 \vert^2 + \vert D \phi_2 \vert^2 \supset \frac{e}{2 R^2} \left( - f^2 + g^2 \right) \partial^\mu \beta \, A_\mu + \frac{e^2}{8 R^2} \left( f^2 + g^2 \right) A_\mu A^\mu
\label{coupling1U1}
\end{equation}
so that in general a linear term in $A_\mu$ is present even in the unitary gauge.

We decompose the scalar fields into background values plus fluctuations, 
\begin{equation}
\phi_i = \langle \phi_i \rangle + \delta \phi_i
\end{equation}
and we restrict our attention to the flat direction background
\begin{eqnarray}
\langle f \rangle = \langle g \rangle \equiv \frac{F}{\sqrt{2}} \;\;\;,\;\;\; \langle \beta \rangle \equiv \Sigma 
\;\; &\Rightarrow& \;\;  \langle \phi_1 \rangle = \langle \phi_2 \rangle = \frac{F}{2 \, R} \, {\rm e}^{i \, \Sigma} 
\end{eqnarray}
(for which the D-term potential vanishes). Since the amplitudes of the two complex fields are equal, the linear term in $A_\mu$ in~(\ref{coupling1U1}) vanishes on this background, so that we can consistently set $\langle A_\mu \rangle = 0 \,$.

From these background assignments, we get the background action
\begin{equation}
S_{\rm bck} = \frac{1}{2} \int d^4 x    \left[ {F'}^2 + F^2 {\Sigma'}^2 - \left(m^2 R^2-\frac{R''}{R} \right)F^2 - \frac{\lambda}{4}F^4 \cos (4\Sigma) \right]
\label{bckacu1}
\end{equation}
from which the first two equations in (\ref{scalarbackground2}) follow. 

We next study the linearized theory for the fluctuations. This amounts to expanding the total action to quadratic order in the fluctuations $\delta \phi_i$ and in $A_\mu$ (since the gauge field has vanishing expectation value, it is treated as a fluctuation). The perturbations of the scalar fields contain three real modes in the unitary gauge. It is convenient to redefine them as
\begin{equation}
\delta f \equiv \frac{r + \delta_H}{\sqrt{2}} \;\;\;,\;\;\;
\delta g \equiv \frac{r - \delta_H}{\sqrt{2}} \;\;\;,\;\;\;
\delta \beta \equiv \frac{\sigma}{F}
\label{pert1u1}
\end{equation}
in terms of which
\begin{eqnarray}
\delta \phi_1 + \delta \phi_2 &=& \left( r + i \, \sigma \right) \, \frac{{\rm e}^{i \, \Sigma}}{R} = \frac{1}{R} \left( \cos \Sigma \, r - \sin \Sigma \, \sigma \right) + \frac{i}{R} \left( \sin \Sigma \, r + \cos \Sigma \, \sigma \right) \equiv \frac{1}{R} \left( \delta_r + i \, \delta_i \right) \nonumber\\
\delta \phi_1 - \delta \phi_2 &=& \delta_H \, \frac{{\rm e}^{i \Sigma}}{R}
\label{rota}
\end{eqnarray}
It is also convenient to decompose the spatial components of the vector field in a transverse plus longitudinal part
\begin{equation}
A_i = A_i^T + \partial_i L
\end{equation}
where $\partial_i A_i^T = 0 \,$.

As we now show, these combinations have an immediate interpretation. Indeed, the quadratic action for the fluctuations splits in three separate pieces
\begin{equation}
S^{(2)} = S_{\rm flat} \left[ \delta_r ,\, \delta_i \right] + S_{\perp} \left[ A_i^T \right] + 
S_{\parallel \rm Higgs }  \left[ \delta_H ,\, L ,\, A_0 \right]
\end{equation}

The modes $\delta_r ,\, \delta_i$ can be interpreted as the real and imaginary excitations of the flat direction. Their quadratic action is
\begin{eqnarray}
S_{\rm flat} = \frac{1}{2} \int d^4 x \left[  \left( \partial_\mu \delta_r \right)^2 + \left( \partial_\mu \delta_i \right)^2 - \left( \delta_r ,\, \delta_i \right) \, \left( \begin{array}{cc}
m^2 R^2 - \frac{R''}{R} + \frac{3 \, \lambda}{2} \, F^2 \, \cos \left( 2 \Sigma \right) &
- \frac{3 \, \lambda}{2} \, F^2 \, \sin \left( 2 \Sigma \right) \\
- \frac{3 \, \lambda}{2} \, F^2 \, \sin \left( 2 \Sigma \right) &
m^2 R^2 - \frac{R''}{R} - \frac{3 \, \lambda}{2} \, F^2 \, \cos \left( 2 \Sigma \right) 
\end{array} \right) \,
\left( \begin{array}{c} \delta_r \\ \delta_i \end{array} \right)
\right] \nonumber\\
\label{fd}
\end{eqnarray}

The quadratic action for the transverse vector modes gets its contributions from the kinetic gauge term $-F^2/4$ and from the second term in~(\ref{coupling1U1}). We find
\begin{equation}
S_{\perp} = \frac{1}{2} \int d^4x \left [ A_i^{T'}A_i^{T'}- (\partial_i A_j^T)\,(\partial_i A_j^T) - \frac{e^2\,F^2}{4} A_i^T\,A_i^T\right]\,.
\label{Sperp}
\end{equation}
where prime denotes (conformal) time derivative.

The remaining fluctuations enter in the quadratic action
\begin{eqnarray}
S_{\parallel \rm Higgs } &=& \frac{1}{2}\int d^4x \Bigg[\delta_H'\,\delta_H'  - (\partial_i \delta_H) \,(\partial_i \delta_H) + (\partial_i L') \,(\partial_i L')+ (\partial_i A_0)\, (\partial_i A_0)
-2\,(\partial_i A_0)\,(\partial_i L')  \nonumber\\
&& \quad\quad\quad
- 2\,e\,F\,\Sigma'\,A_0\,\delta_H -\frac{e^2\,F^2}{4} (\partial_i L)\,(\partial_i L)+\frac{e^2\,F^2}{4} A_0\,A_0
\nonumber\\
&& \quad\quad\quad\left.
-\left(\frac{e^2\,F^2}{4}+m^2 \,R^2-\frac{R''}{R} -\Sigma'^2 -\frac{\lambda}{2} F^2 \cos(4\Sigma)\right)\delta_H \,\delta_H\right]\,.
\label{az1U1LH0}
\end{eqnarray}
The second line is obtained from the coupling term~(\ref{coupling1U1}). We see that the only linear term in the gauge field couples $A_0$ to the mode $\delta_H$. We also see that the mode $A_0$ is nondynamical, and can therefore be integrated out. After Fourier transforming the spatial coordinates, the equation of motion for $A_0$ gives
\begin{equation}
A_0 = 4\,\left(\frac{e\,F\,\Sigma'\,\delta_H+ k^2\,L'}{4\,k^2+e^2\,F^2}\right)\,.
\label{A0sol1U1}
\end{equation}

We can substitute this solution back into the Fourier transform of (\ref{az1U1LH0}). In this way we obtain the action for the two dynamical modes $\delta_H$ and $L \,$. The field $L$ is not canonically normalized. The canonical variable is
\begin{equation}
 A_\parallel \equiv \frac{e\,k\,F}{\sqrt{e^2\,F^2+4\,k^2}} \,L
\end{equation}
In matrix notation, the action takes the form
\begin{equation}
 S_{\parallel \rm Higgs } = \frac{1}{2}\int d\eta \,d^3k \left(\Delta^{\prime\dagger} \Delta' + \Delta^{\prime \dagger} K \Delta - \Delta^\dagger K \Delta^\prime -\Delta^\dagger \Omega^2 \Delta\right) \;\;\;,\;\;\;
 \Delta \equiv \left( \begin{array}{c} \delta_H \\ A_\parallel \end{array} \right)
\label{az1U1LH0s}
 \end{equation}
where the matrix $K$ is antisymmetric, $\Omega^2$ is symmetric, and
\begin{eqnarray}
K_{12} &=& \frac{2\,k\,\Sigma'}{\sqrt{e^2\,F^2+4\,k^2}}\,,
\nonumber\\
(\Omega^2)_{11} &=& \frac{e^2\,F^2}{4} + k^2 + m^2\,R^2-\frac{R''}{R}- \frac{\lambda F^2}{2} \cos(4\,\Sigma) +\left(\frac{3\,e^2\,F^2 - 4\,k^2}{e^2\,F^2+4\,k^2}\right)\Sigma'^2 \,, \nonumber\\
(\Omega^2)_{22} &=& \frac{e^2\,F^2}{4} +\frac{4\,k^2}{e^2\,F^2+4\,k^2}\left[ \frac{e^2\,F^2}{4} + k^2 + m^2\,R^2-\frac{R''}{R}+ \frac{\lambda F^2}{2} \cos(4\,\Sigma)-\Sigma'^2 +\frac{3\,e^2 \, F'^2}{e^2\,F^2+4\,k^2}
\right]\,, \nonumber\\
(\Omega^2)_{12} &=& \frac{k\,F^2}{\sqrt{e^2\,F^2+4\,k^2}} \left[\left(\frac{6\,e^2\,\Sigma'}{e^2\,F^2+4\,k^2}\right)\,\frac{F'}{F} -\lambda \,\sin(4\,\Sigma)\right]
\end{eqnarray}

To eliminate the ``mixed kinetic terms'' (those proportional to $K$) we perform a field redefinition 
$\Psi \equiv {\cal R} \, \Delta$, where ${\cal R}$ is orthogonal (so that the quadratic kinetic term remains canonical). Under such a transformation
\begin{equation}
\frac{1}{2} \Psi^{\prime\dagger} \Psi' = \frac{1}{2} \left( \Delta^{\prime\dagger} \, \Delta'
+ \Delta^{\prime\dagger} \, R^T \, R' \, \Delta - \Delta^\dagger R^T \, R' \Delta' + \Delta^\dagger \, R^{\prime T} \, R' \, \Delta \right)
\end{equation}
where in the third term we have used the fact that $R^T \, R'$ is antisymmetric (due to the orthogonality of $R$). If we now identify
\begin{equation}
\mathcal{R}^T \, \mathcal{R}' = K
\label{defR}
\end{equation}
which gives $K^T \, K = {\mathcal{R}'}^T \, \mathcal{R}'$, the action (\ref{az1U1LH0}) becomes
\begin{equation}
S_{\parallel \rm Higgs } = \frac{1}{2} \int d\eta \,d^3k \, \left[ \Psi^{\prime \dagger} \, \Psi' - \Psi^\dagger \, \mathcal{R} \left( \Omega^2 + K^T K  \right) \mathcal{R}^T \, \Psi \right]
\end{equation}

To proceed, we need to diagonalize the frequency matrix $\mathcal{R} \left( \Omega^2 + K^T K  \right) \mathcal{R}^T$:
\begin{equation}
C^T \, \mathcal{R} \left( \Omega^2 + K^T K  \right) \mathcal{R}^T \, C = \omega^2_{\rm diagonal} \;\;\rightarrow\;\;
\Gamma = C^T \, C'
\label{Cnoneed}
\end{equation}
As outlined in Section \ref{sec:decay}, both the eigenfrequencies and the matrix $\Gamma$ are necessary for computing the production of quanta from the evolving flat direction. The matrix $\mathcal{R}$ is specified by the condition (\ref{defR}) and by the orthogonality requirement. However, the explicit knowledge of $\mathcal{R}$ is not necessary to solve the eigenvalue problem (\ref{Cnoneed}). Firstly, since $\mathcal{R}$ is orthogonal, the eigenvalues of $\mathcal{R} \left( \Omega^2 + K^T K  \right) \mathcal{R}^T$ coincide with those of $\Omega^2 + K^T K$. Secondly, if we simply diagonalize the matrix $\Omega^2 + K^T K$,
\begin{equation}
\xi^T \left( \Omega^2 + K^T K  \right) \xi = {\rm diagonal}
\end{equation}
we can write
\begin{equation}
C = {\cal R} \, \xi \;\;\;\Rightarrow\;\;\; \Gamma = C^T \, C' = \xi^T \, \xi' + \xi^T \, K \, \xi
\label{gam1U1}
\end{equation}
where the condition~(\ref{defR}) has been used. Therefore, we simply need to know the eigenvalues and eigenvectors of $\Omega^2 + K^T K$.

The problem is actually simplified by noting that the off diagonal terms are suppressed in the phenomenologically relevant cases, namely when the conditions~(\ref{seriesapproximation1}) are met.
Disregarding the terms proportional to $\lambda$ and $R''$ (which become negligible as the flat directions start evolving), we find
\begin{equation}
\Omega^2+K^T K = \left(\begin{array}{cc}
 \frac{e^2\,F^2}{4} + \left[ k^2+m^2\,R^2+3\,\Sigma'^2\right] + \mathcal{O}(\epsilon^4) & \mathcal{O}(\epsilon^3)\\
\mathcal{O}(\epsilon^3) & \frac{e^2\,F^2}{4} + \left[ k^2\right] + \mathcal{O}(\epsilon^4)
\end{array}\right)\,,
\end{equation}
where the terms in square parenthesis are of order $\epsilon^2 \,$. The diagonal elements (the eigenfrequencies of the system) vary adiabatically, $\omega' / \omega^2 = \mathcal{O} \left( \epsilon \right) \,$. 

We can also expand the matrix $\Gamma$ obtained in~(\ref{gam1U1}),
\begin{equation}
\Gamma = \left(\begin{array}{cc} 
0&\Gamma_{12}\\-\Gamma_{12}&0
\end{array}\right) \;\;\;,\;\;\;
\Gamma_{12} =
\frac{4\,k\,m^2\,R^2\,\Sigma'}{e\,F\,\left(m^2\,R^2 +3\,\Sigma^{\prime\,2}\right)} \left[2+ \frac{3\,F'}{F\,\left(m^2\,R^2+3\,\Sigma^{\prime\,2}\right)}\left(2\,\frac{F'}{F}+\frac{R'}{R}\right)\right] +\mathcal{O}(\epsilon^4)
\,,
\end{equation}
so that we see that the dominant term is of order $\epsilon^2 \,$. As a consequence, the matrices $I ,\, J$ entering in the equations (\ref{bogequations}) for the particle production are
\begin{equation}
I = \left(\begin{array}{cc} 
0&\mathcal{O}(\epsilon^2)\\\mathcal{O}(\epsilon^2)&0
\end{array}\right) \;\;\;,\;\;\;
J = \left(\begin{array}{cc} 
0&\mathcal{O}(\epsilon^4)\\\mathcal{O}(\epsilon^4)&0
\end{array}\right) \,.
\end{equation}

\section{Two U(1) Flat Directions}
\label{app:twoU1}

Now consider the two flat directions composed of four scalar fields with the potential~(\ref{pot5}).  This potential can accommodate two flat directions, characterized by the background assignments
\begin{equation}
\langle\phi_1\rangle = \langle\phi_2\rangle = \frac{F}{2 R}e^{i\Sigma} \;\;\;,\;\;\;
\langle\phi_3\rangle = \langle\phi_4\rangle = \frac{G}{2 R}e^{i\tilde\Sigma} \;\;\;,\;\;\;
\langle A_\mu \rangle = 0
\label{vevs2U1}
\end{equation}

We parametrize the fluctuations as follows
\begin{eqnarray}
 \phi_1 = \frac{F+r+\delta_H}{2\,R}{\rm e}^{i\,(\Sigma+\sigma/F+\alpha)} \quad\,&,&\quad
 \phi_2 = \frac{F+r-\delta_H}{2\,R}{\rm e}^{i\,(\Sigma+\sigma/F-\alpha)}\,,\nonumber\\
 \phi_3 = \frac{G+\tilde{r}+\tilde{\delta}_H}{2\,R}{\rm e}^{i\,(\tilde{\Sigma}+\tilde{\sigma}/G+\tilde{\alpha})} \quad\,&,&\quad
 \phi_4 = \frac{G+\tilde{r}-\tilde{\delta}_H}{2\,R}{\rm e}^{i\,(\tilde{\Sigma}+\tilde{\sigma}/G-\tilde{\alpha})}\,.
\label{deco2u1}
\end{eqnarray}
where all the fields are real. The scalar fields $\phi_1$ and $\phi_3$ have positive ($+1$) U(1) charge, while $\phi_2$ and $\phi_4$ have negative ($-1$) charge. For this reason, the only linear combination of the phases in (\ref{deco2u1}) which is gauge variant is $\alpha+\tilde{\alpha}$ which we set to zero in the unitary gauge. We also define $a = (\alpha-\tilde{\alpha})/2$. The background value of the latter combination is forced to zero by the equations of motion. 

Calculating the background action yields two copies of (\ref{bckacu1}). As in the single flat direction  case, we decompose the transverse and longitudinal parts of the vector field as:
\begin{equation}
A_i = A_i^T +\partial_i L\,,
\end{equation}
where $\partial_iA_i^T = 0$. At the linearized level, the transverse part of the vector field forms a decoupled system with action
\begin{equation}
S_{\perp} = \frac{1}{2} \int d^4x \left [ A_i^{T'}A_i^{T'}- (\partial_i A_j^T)\,(\partial_i A_j^T) - \frac{e^2\,(F^2+G^2)}{4} A_i^T\,A_i^T\right]\,.
\end{equation}

As in the previous case, the two combinations
\begin{equation}
\delta \phi_1 + \delta \phi_2 = \left( r + i \, \sigma \right) \, \frac{{\rm e}^{i \, \Sigma}}{R} \equiv \frac{1}{R} \left( \delta_r + i \, \delta_i \right) 
\end{equation}
are the real and imaginary excitations of the first flat direction. The quadratic action for these two modes is identical to (\ref{fd}). Analogously, the two combinations
\begin{equation}
\delta \phi_3 + \delta \phi_4 = \left( {\tilde r} + i \, {\tilde \sigma} \right) \, \frac{{\rm e}^{i \, {\tilde \Sigma}}}{R} \equiv \frac{1}{R} \left( {\tilde \delta}_r + i \, {\tilde \delta}_i \right) 
\end{equation}
are the real and imaginary excitations of the second direction. Their quadratic actions is formally of the type (\ref{fd}) with all the quantities referring to the first direction substituted by the analogous ones of the second direction.

The remaining $5$ modes are coupled into the quadratic action
\begin{eqnarray}
S_{\rm coupled} &=& \frac{1}{2}\int d^4x \Bigg[(\partial_\mu \delta_H) \, (\partial^\mu \delta_H)  +(\partial_\mu \tilde{\delta}_H) \, (\partial^\mu \tilde{\delta}_H)+ (\partial_i L') \,(\partial_i L')+ (\partial_i A_0)\, (\partial_i A_0)
\nonumber\\
&& \quad\quad\quad
+(F^2+G^2)(\partial_\mu a)\,(\partial^\mu a)-\frac{e^2\,(F^2 +G^2)}{4} (\partial_i L)\,(\partial_i L)
\nonumber\\
&& \quad\quad\quad
-2\,(\partial_i A_0)\,(\partial_i L') 
- 2\,e\,\left(F\,\Sigma'\,\delta_H +G\,\tilde{\Sigma}'\,\tilde{\delta}_H \right)A_0
\nonumber\\
&& \quad\quad\quad-e\,\left(F^2-G^2\right)\,\left[A_0\,a' -(\partial_i L)\,(\partial_i a)\right] +4\,\left( F\,\Sigma'\,\delta_H- G\,\tilde{\Sigma}'\,\tilde{\delta}_H\right)a'
\nonumber\\
&& \quad\quad\quad
+\frac{e^2\,(F^2+G^2)}{4} A_0\,A_0 -\frac{e^2}{4} \left(F\,\delta_H+G\,\tilde{\delta}_H\right)^2
\nonumber\\
&& \quad\quad\quad
-\left(m^2 \,R^2-\frac{R''}{R} -\Sigma'^2 -\frac{\lambda}{2} F^2 \cos(4\Sigma)\right)\delta_H \,\delta_H
\nonumber\\
&& \quad\quad\quad\left.
-\left(\tilde{m}^2 \,R^2-\frac{R''}{R} -\tilde{\Sigma}'^2 -\frac{\tilde{\lambda}}{2} G^2 \cos(4\tilde{\Sigma})\right)\tilde{\delta}_H \,\tilde{\delta}_H\right]\,.
\end{eqnarray}

To proceed, we Fourier transform this action with respect to the spatial coordinates, and we integrate out the nondynamical field $A_0 \,$. We find
\begin{equation}
A_0 = \frac{1}{k^2 + \frac{e^2}{4}\left(F^2+G^2\right)} \left[\frac{e}{2}\,\left(F^2-G^2\right) \,a' +k^2\,L' + e\,\left( F\,\Sigma'\,\delta_H +G\,\tilde{\Sigma}'\,\tilde{\delta}_H\right)\right]\,.
\label{A0sol2u1}
\end{equation}
We then insert the solution (\ref{A0sol2u1}) back into the Fourier transformed action. The kinetic term of the resulting action is
\begin{eqnarray}
S_{\rm kin} &=& \frac{1}{2} \int d \eta \, d^3k \,\left[ \vert\delta_H^{\prime}\vert^2 +\vert\tilde{\delta}_H^{\prime}\vert^2
+\frac{e^2\,k^2\,\left(F^2+G^2\right)}{e^2\,\left(F^2+G^2\right)+4\,k^2}\,\vert L'\vert^2 + \frac{4\left[e^2\,F^2\,G^2 +k^2\left(F^2+G^2\right)\right]}{e^2\,\left(F^2+G^2\right)+4\,k^2}\,\vert a'\vert^2\right. \nonumber\\
&&\quad\quad\quad\quad\quad\quad
\left. -\frac{2\,e\,k^2\left(F^2-G^2\right)}{e^2\,\left(F^2+G^2\right)+4\,k^2}\,\left( L^{\star\prime}\,a'+L'\,a^{\star\prime}\right) 
\right]
\end{eqnarray}
We canonically normalize the $a$ and $L$ modes through a series of redefinitions. We first define
\begin{equation}
{\hat L} \equiv \frac{e\,k\,\sqrt{F^2+G^2}}{\sqrt{e^2\,(F^2+G^2)+4\,k^2}} \,L \quad\,,\quad
{\hat a} \equiv 2\,\sqrt{\frac{e^2\,F^2\,G^2+k^2\,(F^2+G^2)}{e^2\,(F^2+G^2)+4\,k^2}}\,a\,,
\end{equation}
in terms of which the kinetic term reads
\begin{equation}
S_{\rm kin} = \frac{1}{2} \int d \eta \, d^3k \left[ \vert \delta_H'\vert^{2} +\vert\tilde{\delta}_H'\vert^2 +\vert {\hat L}'\vert^2 + \vert {\hat a}'\vert ^2 -\frac{k\,\left(F^2-G^2\right)}{\sqrt{\left(F^2+G^2\right)\left[e^2\,F^2\,G^2 +k^2\left(F^2+G^2\right)\right]}} \,\left( {\hat L}^{\star \prime}\,{\hat a}' +{\hat L}' {\hat a}^{\star \prime}\right) \right]
\end{equation}
The mixed double derivative is eliminated by 
\begin{equation}
{\hat L}_1 \equiv \frac{{\hat L} - {\hat a}}{\sqrt{2}} \;\;\;,\;\;\;
{\hat L}_2 \equiv \frac{{\hat L} + {\hat a}}{\sqrt{2}}
\end{equation}
Finally, the canonical variables are
\begin{eqnarray}
L_1 \equiv \frac{
\left[ \sqrt{\left(F^2+G^2\right) \left(e^2\,F^2\,G^2 +k^2 (F^2+G^2)\right)}+k\,\left(F^2-G^2\right)\right]^{1/2}
}{
\left[ \left(F^2+G^2\right) \left(e^2\,F^2\,G^2 +k^2 (F^2+G^2)\right)\right]^{1/4}
}\:{\hat L}_1 \nonumber\\
L_2 \equiv \frac{
\left[ \sqrt{\left(F^2+G^2\right) \left(e^2\,F^2\,G^2 +k^2 (F^2+G^2)\right)}-k\,\left(F^2-G^2\right)\right]^{1/2}
}{
\left[ \left(F^2+G^2\right) \left(e^2\,F^2\,G^2 +k^2 (F^2+G^2)\right)\right]^{1/4}
}\: {\hat L}_2 
\end{eqnarray}

In terms of these fields, the action is formally of the type
\begin{equation}
S_{\rm coupled} = \frac{1}{2}\int d\eta \,d^3k \left(\Delta^{\prime\dagger} \Delta' + \Delta^{\prime \dagger} K \Delta - \Delta^\dagger K \Delta^\prime -\Delta^\dagger \Omega^2 \Delta\right)  
 \;\;\;,\;\;\; \Delta \equiv \left( \begin{array}{c} \delta_H \\ {\tilde \delta}_H \\ L_1 \\ L_2 \end{array} \right)
\label{az2U1HHLL}
 \end{equation}
where the matrices $\Omega$ and $K$ are real and, respectively, symmetric and anti-symmetric.

The exact expressions for these matrices are rather involved. However, we can present them as an expansion series in $\epsilon$ (defined in eq.~(\ref{seriesapproximation}); as in the single flat direction 
case, we can neglect the terms proportional to $\lambda ,\, {\tilde \lambda}$ and $R''$). We find
\begin{eqnarray}
K_{12} &=& 0
\,,
\nonumber\\
K_{13} &=& \left\{ \frac{G\,\Sigma'}{\sqrt{2} \,\sqrt{F^2+G^2}} \right\} + \left[ \frac{k\,\Sigma'}{2\,\sqrt{2}\,e\,F} \,\left(\frac{3\,F^2+G^2}{F^2+G^2}\right) \right] +\mathcal{O}(\epsilon^3)
\,,
\nonumber\\
K_{14} &=& \left\{ -\frac{G\,\Sigma'}{\sqrt{2} \,\sqrt{F^2+G^2}} \right\} + \left[ \frac{k\,\Sigma'}{2\,\sqrt{2}\,e\,F} \,\left(\frac{3\,F^2+G^2}{F^2+G^2}\right) \right] +\mathcal{O}(\epsilon^3)
\,,
\nonumber\\
K_{23} &=& \left\{ -\frac{F\,\tilde{\Sigma}'}{\sqrt{2} \,\sqrt{F^2+G^2}} \right\} + \left[ \frac{k\,\tilde{\Sigma}'}{2\,\sqrt{2}\,e\,G} \,\left(\frac{F^2+3\,G^2}{F^2+G^2}\right)\right] +\mathcal{O}(\epsilon^3)
\,,
\nonumber\\
K_{24} &=& \left\{  \frac{F\,\tilde{\Sigma}'}{\sqrt{2} \,\sqrt{F^2+G^2}} \right\} + \left[ \frac{k\,\tilde{\Sigma}'}{2\,\sqrt{2}\,e\,G} \,\left(\frac{F^2+3\,G^2}{F^2+G^2}\right) \right] +\mathcal{O}(\epsilon^3)
\,,
\nonumber\\
K_{34} &=&  \left[ \frac{k\,\left(F^2-G^2\right)}{2\,e\,F\,G\,\left(F^2+G^2\right)^{3/2}}\left(G^2\,\frac{F'}{F}+F^2\,\frac{G'}{G}\right) \right] +\mathcal{O}(\epsilon^4)
\,, \nonumber
\end{eqnarray}
\begin{eqnarray}
(\Omega^2)_{11} &=& \frac{e^2\,F^2}{4} + \left[k^2+m^2\,R^2 +\left(\frac{3\,F^2-G^2}{F^2+G^2}\right)\Sigma'^2\right] +\mathcal{O}(\epsilon^4)
\,, \nonumber\\
(\Omega^2)_{12} &=& \frac{e^2\,F\,G}{4} +\left[ \frac{4\,F\,G}{F^2+G^2}\,\Sigma'\,\tilde{\Sigma}'\right]+\mathcal{O}(\epsilon^4)
\,, \nonumber\\
(\Omega^2)_{13} &=& \left[ \frac{3\,F^2\,G\,\Sigma'}{\sqrt{2}\,\left(F^2+G^2\right)^{3/2}}\left(\frac{F'}{F}-\frac{G'}{G}\right) \right] +\mathcal{O}(\epsilon^3)
\,, \nonumber\\
(\Omega^2)_{14} &=& \left[ -\frac{3\,F^2\,G\,\Sigma'}{\sqrt{2}\,\left(F^2+G^2\right)^{3/2}}\left(\frac{F'}{F}-\frac{G'}{G}\right) \right] +\mathcal{O}(\epsilon^3)
\,, \nonumber
\end{eqnarray}

\begin{eqnarray}
(\Omega^2)_{22} &=& \frac{e^2\,G^2}{4} + \left[k^2+\tilde{m}^2\,R^2 -\left(\frac{F^2-3\,G^2}{F^2+G^2}\right)\tilde{\Sigma}'^2\right] +\mathcal{O}(\epsilon^4)
\,, \nonumber\\
(\Omega^2)_{23} &=& \left[ \frac{3\,F\,G^2\,\tilde{\Sigma}'}{\sqrt{2}\,\left(F^2+G^2\right)^{3/2}}\left(\frac{F'}{F}-\frac{G'}{G}\right) \right] +\mathcal{O}(\epsilon^3)
\,, \nonumber\\
(\Omega^2)_{24} &=& \left[ -\frac{3\,F\,G^2\,\tilde{\Sigma}'}{\sqrt{2}\,\left(F^2+G^2\right)^{3/2}}\left(\frac{F'}{F}-\frac{G'}{G}\right) \right] +\mathcal{O}(\epsilon^3)
\,, \nonumber\\
(\Omega^2)_{33} &=& \frac{e^2\,(F^2+G^2)}{8}+ \left\{ \frac{e\,k\,(F^2-G^2)\,\sqrt{F^2+G^2}}{8\,F\,G} 
\right\}
\nonumber\\
&&+\left[k^2 +\frac{F^2\left(\tilde{m}^2\,R^2-\tilde{\Sigma}'^2\right)+G^2\left(m^2\,R^2-\Sigma'^2\right)}{2\,\left(F^2+G^2\right)} +\frac{3\,\left(G\,F'-F\,G'\right)^2}{2\,\left(F^2+G^2\right)^2}\right] 
+\mathcal{O}(\epsilon^3)
\,, \nonumber\\
(\Omega^2)_{34} &=& 
\frac{e^2\,(F^2+G^2)}{8}\nonumber\\
&&-
\left[
\frac{k^2\left(F^2-G^2\right)^2}{16\,F^2\,G^2} +\frac{F^2\left(\tilde{m}^2\,R^2-\tilde{\Sigma}'^2\right)+G^2\left(m^2\,R^2-\Sigma'^2\right)}{2\,\left(F^2+G^2\right)} +\frac{3\,\left(G\,F'-F\,G'\right)^2}{2\,\left(F^2+G^2\right)^2}\right]+\mathcal{O}(\epsilon^4)
\,, \nonumber\\
(\Omega^2)_{44} &=& 
\frac{e^2\,(F^2+G^2)}{8}- \left\{ \frac{e\,k\,(F^2-G^2)\,\sqrt{F^2+G^2}}{8\,F\,G}\right\} 
\nonumber\\
&&+
\left[k^2 +\frac{F^2\left(\tilde{m}^2\,R^2-\tilde{\Sigma}'^2\right)+G^2\left(m^2\,R^2-\Sigma'^2\right)}{2\,\left(F^2+G^2\right)} +\frac{3\,\left(G\,F'-F\,G'\right)^2}{2\,\left(F^2+G^2\right)^2}\right] +\mathcal{O}(\epsilon^3) \,,
\end{eqnarray}
In these expressions, terms outside parenthesis, within curly parenthesis, and within square parenthesis are, respectively, of zeroth, first, and second order in $\epsilon \,$.

We then proceed as in the previous Appendix by computing the eigenfrequencies, and the matrix $\Gamma$. The leading expressions for these quantities can be found in the main text.

\section{SU(2) flat directions}

\label{app:SU2}

The linearized computation done in the previous two appendices for the U(1) case can be easily extended to nonabelian flat directions of physical interest.  The reason for this is that the differences due to the nonabelian structure arise only at higher than quadratic order in the gauge fields. Such interactions are neglected in the linearized study of fluctuations, as long as the gauge fields involved do not have any background expectation value. Therefore, in the linearized study, the systems of physical fields around nonabelian directions can be treated as a series of copies (one per generator of the group) of those encountered in the abelian case. We show how this is realized in specific examples involving one or two SU(2) flat directions.

Let us start from the case of a single flat direction. We introduce two SU(2) doublets, with the 
potential~(\ref{pot6}), and with the background values
\begin{equation}
\langle \phi_1 \rangle = \frac{F}{2 \, R} \, {\rm e}^{i \, \Sigma} \, 
\left( \begin{array}{c} 1 \\ 0 \end{array} \right)
\;\;\;,\;\;\;
\langle \phi_2 \rangle = \frac{F}{2 \, R} \, {\rm e}^{i \, \Sigma} \, 
\left( \begin{array}{c} 0 \\ 1 \end{array} \right)
\label{back1su2}
\end{equation}
breaking all the SU(2) symmetries. We also assume that the three gauge fields $A_\mu^a$ have vanishing background expectation value.
There are eight real fluctuations of these two complex doublets. It is convenient to parametrize them as
\begin{eqnarray}
\phi_1 = \frac{F+r +\delta_H}{2\,R}\,{\rm e}^{i\,\Sigma} 
\left(\begin{array}{l}
 1+i\,\left(\frac{\delta_3}{F}+\tilde{\delta}_3\right) \\
\frac{\delta_1+i\,\delta_2}{F} + i\,\left(\tilde{\delta}_1+i\,\tilde{\delta}_2\right)
\end{array}
\right)\;\,,\;
\phi_2 = \frac{F+r -\delta_H}{2\,R}\,{\rm e}^{i\,\Sigma} 
\left(\begin{array}{l}
\frac{\delta_1-i\,\delta_2}{F} + i\,\left(\tilde{\delta}_1-i\,\tilde{\delta}_2\right)\\
 1+i\,\left(\frac{\delta_3}{F}-\tilde{\delta}_3\right) 
\end{array}
\right)\,,\nonumber\\
\label{deco1usu2}
\end{eqnarray}

Under the infinitesimal SU(2) transformations $\phi_i \rightarrow U \, \phi_i$, with $U = 1+ i \,\alpha_a\,\sigma^a/2$ (where $\sigma^a$ are the Pauli matrices), only the modes ${\tilde \delta}_i$ are gauge variant, transforming as 
\begin{equation}
\tilde{\delta}_i \rightarrow \tilde{\delta}_i +\frac{\alpha_i}{2}\,.
\end{equation}
In the following, we fix the gauge completely by setting ${\tilde \delta}_i = 0 \,$.

The background action is the same as in the U(1) case given by eq~(\ref{bckacu1}). For the fluctuations, we redefine the fields $r$ and $\delta_3$ as we did for $r$ and $\sigma$ in the first line of (\ref{rota}). We then see that the action can be split in several pieces, analogous to those found for the U(1) case,
\begin{equation}
S=S_{\parallel {\rm Higgs}} (A^1_0,L^1,\delta_1) +S_{\parallel {\rm Higgs}} (A^2_0,L^2,\delta_2)+S_{\parallel {\rm Higgs}} (A^3_0,L^3,\delta_H)+S_{\rm flat}(\delta_r,\delta_i)+\sum_{a=1}^3\,S_\perp (A^{a\,T}_i)\,,
\end{equation}
where $S_{\rm flat}$ is given in eq. (\ref{fd}), the $S_\perp$ actions are three copies of  (\ref{Sperp}), and the $S_{\parallel {\rm Higgs}}$ actions are three copies of (\ref{az1U1LH0}). This confirms what we argued in the main text, namely that the coupled system for the SU(2) flat direction consists of three copies (one per generator) of that obtained in the U(1) case.

This is true also for multiple flat directions. Consider the case of four complex doublets with the potential~(\ref{pot7}). Specifically, we are interested in the case in which two flat directions are present. 
The doublets $\phi_1$ and $\phi_2$ have background values as in (\ref{back1su2}). The doublets $\phi_3$ and $\phi_4$ have analogous values, with $F$ replaced by $G$ and $\Sigma$ by ${\tilde 
\Sigma}$. Similarly to what was done in (\ref{deco1usu2}),  it is convenient to parametrize the $16$ real fluctuations of these fields as 
\begin{eqnarray}
\phi_1 &=& \frac{F+r +\delta_H}{2\,R}\,{\rm e}^{i\,\Sigma} 
\left(\begin{array}{l}
 1+i\,\left(\frac{\delta_3}{F}+\tilde{\delta}_3\right) \\
\frac{\delta_1+i\,\delta_2}{F} + i\,\left(\tilde{\delta}_1+i\,\tilde{\delta}_2\right)
\end{array}
\right)\;\,,\;
\phi_2 = \frac{F+r -\delta_H}{2\,R}\,{\rm e}^{i\,\Sigma} 
\left(\begin{array}{l}
\frac{\delta_1-i\,\delta_2}{F} + i\,\left(\tilde{\delta}_1-i\,\tilde{\delta}_2\right)\\
 1+i\,\left(\frac{\delta_3}{F}-\tilde{\delta}_3\right) 
\end{array}
\right)\,, \nonumber\\
\phi_3 &=& \frac{G+\tilde{r} +\tilde{\delta}_H}{2\,R}\,{\rm e}^{i\,\tilde{\Sigma}} 
\left(\begin{array}{l}
 1+i\,\left(\frac{\gamma_3}{G}+\tilde{\gamma}_3\right) \\
\frac{\gamma_1+i\,\gamma_2}{G} + i\,\left(\tilde{\gamma}_1+i\,\tilde{\gamma}_2\right)
\end{array}
\right)\;\,,\;
\phi_4 = \frac{G+\tilde{r} -\tilde{\delta}_H}{2\,R}\,{\rm e}^{i\,\tilde{\Sigma}} 
\left(\begin{array}{l}
\frac{\gamma_1-i\,\gamma_2}{G} + i\,\left(\tilde{\gamma}_1-i\,\tilde{\gamma}_2\right)\\
 1+i\,\left(\frac{\gamma_3}{G}-\tilde{\gamma}_3\right) 
\end{array}
\right)\,. \nonumber\\
\label{deco2usu2}
\end{eqnarray}

In this case, the gauge dependent combinations of the perturbations are $\tilde{\delta}_i+\tilde{\gamma}_i$. We fix the gauge completely by setting them to zero, and we define $a_i =(\tilde{\delta}_i - \tilde{\gamma}_i)/2$.

The background action yields two copies of (\ref{bckacu1}). Transforming the fields $(r,\delta_3)$ and $(\tilde{r},\gamma_3)$ as in the first line of (\ref{rota}), we can express the quadratic action in terms of the 
separated actions obtained in the U(1) case,
\begin{eqnarray}
S&=&S_{\rm coupled} (A^1_0,L^1,\delta_1,\gamma_1,a_1) +S_{\rm coupled} (A^2_0,L^2,\delta_2,\gamma_2,a_2)+S_{\rm coupled} (A^3_0,L^3,\delta_H, \tilde{\delta}_H,a_3)\nonumber\\
&&+S_{\rm flat}(\delta_r,\delta_i)+S_{\rm flat}(\tilde{\delta}_r,\tilde{\delta}_i)+\sum_{a=1}^3\,S_\perp (A^{a\,T}_i)\,, 
\end{eqnarray}
Here as well, the coupled system also consists of three copies of the coupled system obtained in U(1) case.

%
%
\section{Equations in program units}
\label{app:eqsprog}

We list here the closed set of equations that we solve numerically in Section \ref{sec:results}. In terms of the dimensionless quantities (\ref{program}), the background equations are
\begin{eqnarray}
&&F_*''+\left(m_*^2 R^2- \frac{R''}{R}-\Sigma'^2 \right) F_* + \frac{\lambda_*}{2} \,F_*^3 \cos\left(4\,\Sigma\right) = 0\nonumber \\
&&\left(F_*^2\,\Sigma'\right)' -\frac{\lambda_*}{2} \,F_*^4 \sin\left(4\,\Sigma\right)=0 \nonumber \\
&&G_*''+\left({\tilde m}_*^2 R^2- \frac{R''}{R}-{\tilde \Sigma}'^2 \right) G_* + \frac{{\tilde \lambda}_*}{2} \,G_*^3 \cos\left(4\,{\tilde \Sigma}\right) = 0\nonumber \\
&&\left(G_*^2\,{\tilde \Sigma}'\right)' -\frac{{\tilde \lambda}_*}{2} \,G_*^4 \sin\left(4\,{\tilde \Sigma}\right)=0 \nonumber \\
&& \frac{R''}{R} + \frac{R'^2}{R^2} = 4 \pi \left\{ \frac{\vert \Phi_0 \vert^2}{M_p^2} \left[
m_*^2 \, F_*^2 + \frac{\lambda_* \, F_*^4}{4 \, R^2} \cos \left( 4 \Sigma \right)
+ {\tilde m}_*^2 \, G_*^2 + \frac{{\tilde \lambda_*} \, G_*^4}{4 \, R^2} \cos \left( 4 {\tilde \Sigma} \right) \right] + R^2 \, \rho_{\psi*} \right\}
\label{bckprog}
\end{eqnarray}
where prime now denotes derivatives with respect to $\eta_* \,$.

Eqs. (\ref{bogequations}) for the produced quanta read
\begin{eqnarray}
& \alpha' = \left(-i \omega_* -I_*\right)\alpha + \left(\frac{\omega_*'}{2\omega_*}-J_*\right)\beta \nonumber \\
& \beta' = \left(i \omega_*-I_*\right) \beta + \left(\frac{\omega_*'}{2\omega_*} -J_*\right)\alpha
\label{bogprog}
\end{eqnarray}
where we have introduced the dimensionless frequencies
\begin{equation}
\omega_* \equiv {\rm diag} \left( \omega_{1*} ,\, \omega_{2*} ,\, \omega_{3*} \right)
\;\;\;,\;\;\; \omega_{i*} \equiv \sqrt{\frac{m_i^2 \, R^2}{e^2 \, \vert \Phi_0 \vert^2} + k_*^2}
\label{freprog}
\end{equation}
with the eigenmasses (\ref{masseigenvalues}), and the dimensionless matrices
\begin{equation} 
I_*,J_* = \frac12 \left(\sqrt{\omega_*} \Gamma_* \frac{1}{\sqrt{\omega_*}} \pm \frac{1}{\sqrt{\omega_*}}\Gamma_* \sqrt{\omega_*} \right) \;\;\;,\;\;\; \Gamma_* \equiv \frac{\Gamma}{e \, \vert \Phi_0 \vert}
\label{ijprog}
\end{equation}
with the $\Gamma$ matrix specified in (\ref{gammamatrix}). The quantities $\omega_{*}$ and $\Gamma_*$ can be immediately written in terms of the dimensionless variables (\ref{program}). One can check that both the values of $e$ and $\vert \Phi_0 \vert$ are rescaled out from these expressions.

%
%
\section{Approximate scaling of the solutions}
\label{app:scaling}

The numerical solutions exhibit an approximate scaling with $\epsilon$, defined after equation 
(\ref{seriesapproximation}). To see this, we rescale the amplitudes and the masses of the flat directions by two different constant factors,
\begin{equation}
\left\{ \Phi_0 ,\, {\tilde\Phi_0} \right\} \rightarrow \left\{ \gamma \, \Phi_0 ,\, \gamma \, {\tilde \Phi_0} \right\} \;\;\;,\;\;\; \left\{ m ,\, {\tilde m} \right\} \rightarrow \left\{ \mu \, m ,\, \mu \, {\tilde m} \right\} \,,
\label{rescalingsapp}
\end{equation}
We already showed (see for instance the discussion after eq.(\ref{eombck})) that, under the rescaling (\ref{rescalingsapp}), a background solution $\Phi \left( t \right) ,\, {\tilde \Phi} \left( t \right)$ is mapped into the solution $\gamma \, \Phi \left( \mu \, t \right) ,\, \gamma \, {\tilde \Phi} \left( \mu \, t \right) \,$. Therefore, 
$\mu$ only affects the timescale governing the dynamics of the flat directions. From this, we can see that, under the rescaling (\ref{rescalingsapp}), each quantity on the left hand side of 
(\ref{seriesapproximation}) is multiplied by $\mu$, while the quantities at the right hand side by $\gamma \,$. Therefore, under this rescaling,
\begin{equation}
\epsilon \rightarrow \left( \frac{\mu}{\gamma} \right) \epsilon
\label{epssca}
\end{equation}

To understand how particle production changes under (\ref{rescalingsapp}), it is useful to write the
corresponding equations in the form (\ref{bogprog}). Particle production takes place at momenta comparable to the flat direction masses. From the definitions (\ref{freprog}), and from eqs. (\ref{masseigenvalues}), we see that
\begin{equation}
\omega_{1*} = \mathcal{O} \left( 1 \right) \;\;,\;\; \omega_{2,3*} = \mathcal{O} \left( \epsilon \right) 
\end{equation}
For this reason we define
\begin{equation}
{\bar \omega}_1 \equiv \omega_{*1} \;\;,\;\; {\bar \omega}_{2,3} \equiv \frac{\omega_{*2,3}}{\epsilon}
\end{equation}
so that all ${\bar \omega}_i$ are of order one. We also note that they do not change under the rescalings (\ref{rescalingsapp}).

Concerning the matrices $I_*$ and $J_*$, we have instead (cf. the definitions (\ref{ijprog}) 
and eqs.~(\ref{gammamatrix}))
\begin{eqnarray}
&& I_{13*} = - I_{31*} = \mathcal{O} \left( \epsilon^{1/2} \right) \;\;\;,\;\;\;
I_{23*} = - I_{32*} = \mathcal{O} \left( \epsilon \right) \nonumber\\
&& J_{13*} = J_{31*} = \mathcal{O} \left( \epsilon^{1/2} \right) \;\;\;,\;\;\;
J_{23*} = J_{32*} = \mathcal{O} \left( \epsilon \right) 
\end{eqnarray}
while all the other elements are of higher order in $\epsilon$ and can be neglected. As for the frequencies, we define 
\begin{equation}
{\bar I}_{13} \equiv \frac{I_{13*}}{\epsilon^{1/2}} \;\;,\;\;
{\bar I}_{23} \equiv \frac{I_{23*}}{\epsilon} \;\;,\;\;
{\bar J}_{13} \equiv \frac{J_{13*}}{\epsilon^{1/2}} \;\;,\;\;
{\bar J}_{23} \equiv \frac{J_{23*}}{\epsilon}
\end{equation}
so that all of these quantities are of order one, and unaffected by the rescalings (\ref{rescalingsapp}).

Finally, we rescale the time variable as ${\bar \eta} \equiv \eta_* / \epsilon \,$; once written this way, the time dependence of the background solutions is unaffected by the rescalings 
(\ref{rescalingsapp}) (namely, $\Phi \left( {\bar \eta} \right) \rightarrow \gamma \, \Phi \left( {\bar \eta} \right)
\;,\; {\tilde \Phi} \left( {\bar \eta} \right) \rightarrow \gamma \, {\tilde \Phi} \left( {\bar \eta} \right) \,$). This guarantees that the three quantities $d \, {\bar \omega}_i / d \, {\bar \eta}$ are of order one, and unaffected by the rescalings (\ref{rescalingsapp}).

In terms of these variable, eqs.~(\ref{bogprog}) read
\begin{eqnarray}
\epsilon \, \alpha_{1j}' &=& - i \, {\bar \omega}_1 \, \alpha_{1j} - \epsilon^{1/2} \left( {\bar I}_{13} \, \alpha_{3j} + {\bar J}_{13} \, \beta_{3j} \right) + \epsilon \, \frac{{\bar \omega}_1'}{2 \, {\bar \omega}_1} \, \beta_{1j} \nonumber\\
\epsilon \, \beta_{1j}' &=& i \, {\bar \omega}_1 \, \beta_{1j} - \epsilon^{1/2} \left( {\bar I}_{13} \, \beta_{3j} + {\bar J}_{13} \, \alpha_{3j} \right) + \epsilon \, \frac{{\bar \omega}_1'}{2 \, {\bar \omega}_1} \, \alpha_{1j} \nonumber\\
\epsilon \, \alpha_{2j}' &=& - \epsilon \left( i \, {\bar \omega}_2 \, \alpha_{2j} + {\bar I}_{23} \, \alpha_{3j} + 
{\bar J}_{23} \, \beta_{3j} - \frac{{\bar \omega}_2'}{2 \, {\bar \omega}_2} \, \beta_{2j} \right) \nonumber\\
\epsilon \, \beta_{2j}' &=& \epsilon \left( i \, {\bar \omega}_2 \, \beta_{2j} - {\bar I}_{23} \, \beta_{3j} - 
{\bar J}_{23} \, \alpha_{3j} + \frac{{\bar \omega}_2'}{2 \, {\bar \omega}_2} \, \alpha_{2j} \right) \nonumber\\
\epsilon \, \alpha_{3j}' &=& \epsilon^{1/2} \left( {\bar I}_{13} \, \alpha_{1j} - {\bar J}_{13} \, \beta_{1j} \right) + \epsilon \left( - i \, {\bar \omega}_3 \, \alpha_{3j} + {\bar I}_{23} \, \alpha_{2j} - {\bar J}_{23} \, \beta_{2j} + \frac{{\bar \omega}_3'}{2 \, {\bar \omega}_3} \, \beta_{3j} \right) \nonumber\\
\epsilon \, \beta_{3j}' &=& \epsilon^{1/2} \left( {\bar I}_{13} \, \beta_{1j} - {\bar J}_{13} \, \alpha_{1j} \right) + \epsilon \left( i \, {\bar \omega}_3 \, \beta_{3j} + {\bar I}_{23} \, \beta_{2j} - {\bar J}_{23} \, \alpha_{2j} + \frac{{\bar \omega}_3'}{2 \, {\bar \omega}_3} \, \alpha_{3j} \right) 
\label{eqscaling}
\end{eqnarray}
where $j=1,2,3 \,$, and where now prime denotes derivative with respect to ${\bar \eta} \,$. As we mentioned, all of the quantities ${\bar \omega} ,\, {\bar I} ,\, {\bar J} ,\, {\bar \omega'}$ in these equations are of order one, and do not change under the rescalings (\ref{rescalingsapp}). In contrast, $\epsilon$ is small and changes according to (\ref{epssca}). Due to this hierarchy, we can neglect the last term in the first two equations (\ref{eqscaling}). The numerical solutions also show that, as soon as the quanta are produced in an appreciable number, $\vert \epsilon \, \alpha_{1j}' \vert \ll \vert {\bar \omega}_1 \, \alpha_{1j} \vert$ and $\vert \epsilon \, \beta_{1j}' \vert \ll \vert {\bar \omega}_1 \, \beta_{1j} \vert$. Therefore, the first two equations of~(\ref{eqscaling}) can be written as
\begin{eqnarray}
0 &\simeq& - i \, {\bar \omega}_1 \, \left( \epsilon^{-1/2} \, \alpha_{1j} \right) - {\bar I}_{13} \, \alpha_{3j} - {\bar J}_{13} \, \beta_{3j}  \nonumber\\
0 &\simeq& i \, {\bar \omega}_1 \, \left( \epsilon^{-1/2} \, \beta_{1j} \right) -  {\bar I}_{13} \, \beta_{3j} - {\bar J}_{13} \, \alpha_{3j}  
\label{eqscaling2}
\end{eqnarray}
while the remaining four equations can be cast in the form
\begin{eqnarray}
\alpha_{2j}' &=& -  i \, {\bar \omega}_2 \, \alpha_{2j} - {\bar I}_{23} \, \alpha_{3j} - 
{\bar J}_{23} \, \beta_{3j} + \frac{{\bar \omega}_2'}{2 \, {\bar \omega}_2} \, \beta_{2j}  \nonumber\\
\beta_{2j}' &=& i \, {\bar \omega}_2 \, \beta_{2j} - {\bar I}_{23} \, \beta_{3j} - 
{\bar J}_{23} \, \alpha_{3j} + \frac{{\bar \omega}_2'}{2 \, {\bar \omega}_2} \, \alpha_{2j} \nonumber\\
\alpha_{3j}' &=& {\bar I}_{13} \, \left( \epsilon^{-1/2} \, \alpha_{1j} \right) - {\bar J}_{13} \, 
\left( \epsilon^{-1/2} \, \beta_{1j} \right) - i \, {\bar \omega}_3 \, \alpha_{3j} + {\bar I}_{23} \, \alpha_{2j} - {\bar J}_{23} \, \beta_{2j} + \frac{{\bar \omega}_3'}{2 \, {\bar \omega}_3} \, \beta_{3j} \nonumber\\
\beta_{3j}' &=& {\bar I}_{13} \, \left( \epsilon^{-1/2} \, \beta_{1j} \right) - {\bar J}_{13} \left( \epsilon^{-1/2} \, \alpha_{1j} \right) + i \, {\bar \omega}_3 \, \beta_{3j} + {\bar I}_{23} \, \beta_{2j} - {\bar J}_{23} \, \alpha_{2j} + \frac{{\bar \omega}_3'}{2 \, {\bar \omega}_3} \, \alpha_{3j}
\label{eqscaling4}
\end{eqnarray}

From a quick inspection of eqs.~(\ref{eqscaling2}) and (\ref{eqscaling4}) we conjecture that, if we
compare the particle production in several cases which differ from each other only on the value of 
$\epsilon$, we will find that
\begin{equation}
\alpha_{1j} , \beta_{1j} \propto \epsilon^{1/2} \;\;\;,\;\;\; \alpha_{2j} ,\, \beta_{2j} ,\, \alpha_{3j} ,\, \beta_{3j} \propto \epsilon^0
\end{equation}
Recalling that the occupation number of the ith mode is, $n_i = \sum_j \beta_{ij}\beta_{ij}^*$, this implies
\begin{equation}
n_1 \propto \epsilon \;\;\;,\;\;\; n_2 ,\, n_3 \propto \epsilon^0
\end{equation}

Equivalently, if we compute the particle production for some given configuration, and we then consider a second configuration related to the first one by the rescalings (\ref{rescalingsapp}), we conjecture that
\begin{eqnarray}
&& \left\{ \alpha_{1j}\;,\;\beta_{1j}\;,\;\alpha_{2j}\;,\;\beta_{2j}\;,\;\alpha_{3j}\;,\;\beta_{3j}\right\} \rightarrow \left\{\sqrt{\frac{\mu}{\gamma}} \, \alpha_{1j}\;,\;\sqrt{\frac{\mu}{\gamma}} \, \beta_{1j}\;,\;\alpha_{2j}\;,\;\beta_{2j}\;,\;\alpha_{3j}\;,\;\beta_{3j} \right\} \nonumber\\
&& \left\{ n_1 \;,\; n_2 \;,\; n_3 \right\} \rightarrow \left\{ \frac{\mu}{\gamma} \, n_1 \;,\; n_2 \;,\; n_3 \right\}
\end{eqnarray}
The numerical computations that we have performed confirm this scaling behavior 
(see Section \ref{sec:results}).
It is not hard to verify that, under the rescalings (\ref{rescalingsapp}), the dimensionless quantities defined in (\ref{program}) scale as
\begin{eqnarray}
&& F_* \rightarrow F_* \;\;,\;\; 
G_* \rightarrow G_* \;\;,\;\; 
m_* \rightarrow \frac{\mu}{\gamma} \, m \;\;,\;\; 
{\tilde m}_* \rightarrow \frac{\mu}{\gamma} \, {\tilde m}_* \;\;,\;\;
\lambda_* \rightarrow \frac{\mu^2}{\gamma^2} \, \lambda_* \;\;,\;\;
{\tilde \lambda}_* \rightarrow \frac{\mu^2}{\gamma^2} \, {\tilde \lambda}_* \nonumber\\
&& \frac{d}{d \, \eta_*} \left[ {\rm background} \right] \rightarrow \frac{\mu}{\gamma} 
\, \frac{d}{d \, \eta_*} \left[ {\rm background} \right] \;\;,\;\;
k_* \rightarrow \frac{\mu}{\gamma} \, k_* \;\;,\;\;
\omega_{i*} \, n_i \rightarrow \frac{\mu}{\gamma} \, \omega_{i*} \, n_i
\end{eqnarray}
where ``background'' stands for any background quantity. Therefore, the ratio between the energy density of the produced quanta and that of the flat directions, given in eq.~(\ref{energydensityratio}), scales as
\begin{equation}
r_{\rm prod} \rightarrow \frac{\mu^2}{\gamma^2} \, r_{\rm prod}
\end{equation}
Equivalently, we can say that $r_{\rm prod} \propto \epsilon^2 \,$.

%


%
\newpage

\end{document}